\documentclass[11pt]{article}

\usepackage{doublespace}
\usepackage{epsfig}
\usepackage{amssymb}
\usepackage{verbatim}
\usepackage{delarray}
\usepackage{graphicx}
\usepackage{epic}
\usepackage{amsthm}
\usepackage{amstext}

\newcommand{\bl}{\left(}
\newcommand{\br}{\right)}
\newcommand{\re}[1]{\Re\{ {#1} \}}

\newcommand{\tr}{\mbox{Tr}}

\newcommand{\lra}{\leftrightarrow}
\newcommand{\inner}[2]{\langle{#1},{#2}\rangle}

\newcommand{\hmec}{\widehat{F}_c}
\newcommand{\hmc}{$\widehat{F}_c \,$}
\newcommand{\hmeu}{\widehat{F}_u}
\newcommand{\hmu}{$\widehat{F}_u \,$}

\newcommand{\HH}{{\mathcal{H}}}

\newcommand{\V}{{\mathcal{V}}}

\newcommand{\U}{{\mathcal{U}}}
\newcommand{\SSS}{{\mathcal{S}}}
\newcommand{\G}{{\mathcal{G}}}
\newcommand{\C}{{\mathcal{C}}}
\newcommand{\FF}{{\mathcal{F}}}
\newcommand{\Z}{{\mathbb{Z}}}
\newcommand{\CCC}{{\mathbb{C}}}

\newcommand{\ie}{{\em i.e., }}
\newcommand{\eg}{{\em e.g., }}

\newcommand{\eqr}[1]{(\ref{#1})}

\newcommand{\ket}[1]{{|#1\rangle}}
\newcommand{\bra}[1]{{\langle#1|}}
\newcommand{\braket}[2]{{\langle#1|#2\rangle}}

\newtheorem{theorem}{Theorem}

\title{Optimal Tight Frames and Quantum Measurement}
\author{Yonina C. Eldar\footnote{Research Laboratory of Electronics,
Massachusetts Institute of Technology, Room 36-615,
Cambridge, MA 02139. E-mail: yonina@mit.edu.}
\hspace*{0.05in} and G. David Forney, Jr.\footnote{Laboratory for
Information and Decision Systems, Massachusetts Institute of
Technology, Cambridge, MA 02139. E-mail: forneyd@mediaone.net.}}

\begin{document} 

\maketitle

\begin{abstract}

Tight frames and rank-one quantum measurements are shown to be
intimately related.  In fact, the family of normalized tight frames
for the space in which a quantum mechanical system lies is precisely
the family of rank-one generalized quantum measurements (POVMs) on
that space.  Using this relationship, frame-theoretical analogues of
various quantum-mechanical concepts and results are developed.

The analogue of a least-squares quantum measurement is a tight frame
that is closest in a least-squares sense to a given set of vectors.
The least-squares tight frame is found for both the case in which the
scaling of the frame is specified (constrained least-squares frame
(CLSF)) and the case in which the scaling is free (unconstrained
least-squares frame (ULSF)).  The well-known canonical frame is shown
to be proportional to the ULSF and to coincide with the CLSF with a
certain scaling.

Finally, the canonical frame vectors corresponding to a geometrically
uniform vector set are shown to be geometrically uniform and to have
the same symmetries as the original vector set.

\end{abstract}


\section{Introduction}

Frames are generalizations of bases which lead to redundant signal
expansions \cite{DS52,Y80}.  A frame for a Hilbert space
$\U$ is a set of not necessarily linearly independent vectors that
spans $\U$ and has some additional
properties. Frames were first introduced by Duffin and Schaeffer \cite{DS52}
in the
context of nonharmonic Fourier series, and play an important role in the
theory of
nonuniform sampling
\cite{DS52,Y80,B92}. Recent interest in frames has been motivated in
part by their utility in analyzing wavelet expansions \cite{HW89,D90}.

Many efforts have been made to construct
bases with specified properties. Since the conditions on bases are quite
stringent, in many applications it is hard to find ``good" bases. The
conditions
on frame vectors are usually not as stringent, allowing for
increased flexibility in their design \cite{HW89,A95}.  For example, frame
expansions admit signal representations that are localized in both time and
frequency \cite{D90}, as well as sparse representations \cite{AHSE01}.

Frame expansions have many other desirable properties. The
coefficients may be computed with less precision than the coefficients
in a basis expansion for a given desired reconstruction precision
\cite{D90}; the effect of additive noise on the coefficients on the
reconstructed signal is reduced in comparison with a basis expansion
\cite{D90,D92,M92,E012}; and the coefficients are more robust to
quantization degradations \cite{CV96,GVT98}.  Recently, frames have
been applied to the development of modern uniform and nonuniform
sampling techniques \cite{AG01}, to various detection problems
\cite{EO01,EC01}, and to the analysis and design of packet-based
communication systems \cite{GKV98}.

A \emph{tight frame} is a special case of a frame for which the
reconstruction
formula is particularly simple.  As we show in
Section~\ref{sec:frames}, a tight frame expansion of a signal is
reminiscent of
an orthogonal basis expansion, even though the frame vectors in the
expansion are
linearly dependent. Tight frames are
particularly popular, and will be the focus of this paper.

Frame-like expansions have been developed and used in a wide range of
disciplines. Many connections between frame theory and various signal
processing techniques have been recently discovered and developed. For
example, the
theory of frames has been used to analyze and design oversampled filter
banks
\cite{B97,BHF98} and error correction codes
\cite{F99}.  Wavelet families have been used in quantum mechanics and
many other areas of theoretical physics, particularly in the study of
semiclassical approximations to quantum mechanics \cite{D90}.

In this paper we explore yet another connection between quantum
mechanics and tight frames.  Specifically, we show that the
family of (normalized) tight frames for a subspace $\U$ in which a quantum
mechanical system is known to lie is precisely
the family of possible generalized measurements (POVMs) on $\U$. Exploiting
this
equivalence, we can apply ideas and results derived in the context of
quantum measurement to the theory of frames and \emph{vice versa}.

We begin in Section~\ref{sec:qm} by characterizing quantum
measurements. With each rank-one quantum measurement we associate a
measurement
matrix.  Using the measurement matrix representation, we give a
simple and constructive proof of Neumark's theorem \cite{P90}, which
relates general quantum measurements to orthogonal measurements.  We
then discuss the problem of constructing measurements optimized to
distinguish between a set of non-orthogonal pure quantum states.

We then follow a similar path in Section~\ref{sec:frames} for tight
frames.  We associate a frame matrix with every tight frame, which as
we show has essentially the same properties as a quantum measurement
matrix. Next, we derive an analogue of Neumark's theorem for tight
frames, which expresses tight frame vectors as projections of a set of
orthogonal
vectors in a larger space.  Finally, motivated by the construction of
optimal quantum measurements, we consider the problem of constructing
optimal tight frames for a subspace $\U$ from a given set of vectors
that span $\U$.

The problem of frame design has received relatively little attention
in the frame literature. Typically in applications the frame vectors
are chosen, rather than optimized. Iterative algorithms for
constructing frames that are optimal in some sense are given in
\cite{EAH98}.  Methods for generating frames starting from a given
frame are described in \cite{A95}.

A popular frame construction from a
given set of vectors is the canonical frame \cite{D92,B97,BJ99,JB00},
first proposed in the context of wavelets in \cite{M86}.  The
canonical frame is relatively simple to construct, can be
determined directly from the given vectors, and plays an important
role in wavelet theory \cite{D88,M89,UA93}.  However, no general
optimality properties are known for the canonical frame.

In Section~\ref{sec:optimal} we systematically construct optimal
frames from a given set of vectors.  Motivated by the least-squares
measurement \cite{EF01} derived for quantum detection,
we seek a tight frame consisting of frame vectors that minimize the
sum of the squared norms of the error vectors, where the $i$th error
vector is defined as the difference between the $i$th given vector and
the $i$th frame vector. We consider both the case in which the
scaling of the frame is specified, and the case in which the
scaling is such that the error is minimized.  When the
scaling is specified the optimizing frame is referred to as the
constrained least-squares frame (CLSF), and when the scaling is
not specified the optimizing frame is referred to as the unconstrained
least-squares frame (ULSF).

In Section~\ref{sec:compare} we show that the
canonical frame vectors
are proportional to the ULSF vectors, and that they coincide with the CLSF
vectors with a specific choice of scaling.

An important issue is to what extent frames constructed from a given
set of vectors inherit the properties of the original vector set
\cite{BJ99}. For example, it has been shown that when constructing
normalized Gabor frames from windows satisfying certain decay
conditions using the canonical frame construction, the resulting tight
frame has similar decay properties \cite{BJ99}.  In
Section~\ref{sec:gu} we consider the case in which the original
vectors have a strong symmetry property called geometric uniformity
\cite{F91}. Based on results derived in the context of quantum
detection \cite{EF01} we show that the CLSF vectors and the ULSF
vectors have the same symmetries as the original vectors. This implies that
the
canonical frame vectors associated with a geometrically uniform vector set
are
themselves geometrically uniform.

Before proceeding to the detailed development, in
Section~\ref{sec:prelim} we first
provide an overview of the notation and some mathematical preliminaries.

\section{Preliminaries}
\label{sec:prelim}

In this section we briefly review elements of linear algebra that are common
to
both signal processing and quantum mechanics.  Our main goal is to
characterize
``transjectors" (partial isometries) using the singular value
decomposition (SVD).

\subsection{Hilbert spaces and operators}

In both signal processing and quantum mechanics, the setting we
consider is a 
finite-dimensional subspace $\U$ of a complex Hilbert space $\HH$.
The elements
of $\HH$ are called vectors.
We will often assume for notational convenience that $\HH$ is
finite-dimensional, with $\dim \HH = k$;  then by appropriate choice of
coordinates we can identify $\HH$ with $\CCC^k$.

In signal processing, the elements of $\HH$ are
regarded as column vectors and denoted, \eg by $x \in \HH$.  Then $x^*$
denotes
the row vector which is the conjugate transpose of $x$.  The inner product
of two
vectors is a complex number, denoted, \eg by $\inner{x}{y} = x^* y$.  An
outer
product of two vectors such as $x y^*$ is a rank-one matrix, which as an
operator
takes $z \in \HH$ to $x y^* z = \inner{y}{z} x \in \HH$.

The Dirac bra-ket notation of quantum mechanics expresses such concepts very
nicely.  We believe that the  signal processing community would do
well to master 
it; 
however, recognizing that it is unfamiliar, we do not rely on it in this
paper.  Nonetheless, to assist the reader unfamiliar with this notation
in  reading the quantum literature, we
will give the bra-ket equivalents for various expressions in this section.

In the bra-ket notation, the elements of $\HH$ are ``ket" vectors,
denoted, \eg by $\ket{x} \in \HH$.  The corresponding ``bra" vector
$\bra{x}$ is an element of the dual space $\HH^*$ and may be regarded
as the conjugate transpose of $\ket{x}$.  The inner product of two
vectors is a complex number denoted by $\braket{x}{y}$.  An outer
product of two vectors such as
$\ket{x}\bra{y}$ is a rank-one matrix, which as an operator takes
$\ket{z} \in \HH$ to $\ket{x}\bra{y}\ket{z} = \braket{y}{z} \ket{x}
\in \HH$.

An operator on $\HH$ is a linear transformation $A: \HH \to \HH$.  The
adjoint of
an operator $A$ is the unique operator $A^*$ such that $\inner{x}{Ay} =
\inner{A^*x}{y}$ for all $x,y \in \HH$.  If the elements of $\HH$ are column
vectors, then an operator $A$ is represented by a square matrix, and its
adjoint
is represented by the conjugate transpose $A^*$, since  $\inner{x}{Ay} = x^*
Ay =
(A^* x)^* y = \inner{A^*x}{y}$.

 An operator $A$ is called \emph{Hermitian} if it is self-adjoint;  \ie if
$A^* = A$.

An orthogonal projector $P$ is a Hermitian operator on $\HH$ such that
$P^2 = P$.  Consequently, the eigenvalues of $P$ all equal 0 or 1.  If
$\{u_i\}$ is a set of eigenvectors corresponding to the nonzero
eigenvalues of $P$, then the subspace $\U \subseteq \HH$ spanned by
the set $\{u_i\}$ is the range of $P$, and we write the projector as
$P_{\U}$.  A one-dimensional projector has a single eigenvector $u$
and may be written as the outer product $P_{u} = u u^*$ (or $P_{u} =
\ket{u}\bra{u}$ in bra-ket notation); then $P_{u}$ projects any $x \in
\HH$ into the projection $P_{u}x = \inner{u}{x} u$ (or
$\ket{u}\braket{u}{x}$).  An $r$-dimensional projector $P_{\U}$ may be
written as the sum of $r$ one-dimensional projectors, $P_{\U} = \sum_i
P_{u_i}$, where $\{u_i\}$ is any basis for $\U$.

\subsection{Transjectors (partial isometries)}

Let $F$ be a rank-$r$ matrix whose columns are a set of $n$ vectors
$\varphi_i \in \HH$.  Then $F^*x$ is a vector in $\CCC^n$ whose
components are the inner products $\inner{\varphi_i}{x}$.  In other
words, $F^*$ may be regarded as a linear transformation $F^*: \HH \to
\CCC^n$.  Similarly, $F$ may be regarded as a linear transformation
$F: \CCC^n \to \HH$.

It is well known in signal processing (but not as well known in
quantum mechanics) that any such matrix $F$ has an SVD $F = U\Sigma
V^*$, where $U$ is a unitary matrix whose columns $\{u_i \in \HH\}$
are the eigenvectors of the Hermitian operator $T = FF^*$, $V$ is a
unitary matrix whose columns $\{v_i \in \CCC^n\}$ are the eigenvectors
of the Hermitian matrix $S = F^* F$ (the Gram matrix of inner
products), and $\Sigma$ is a positive real diagonal matrix whose $r$
nonzero values $\sigma_i$, called the \emph{singular values} of $F$,
are the positive square roots of the nonzero eigenvalues of either $S$
or $T$.  Thus we may write $F = \sum_i \sigma_i u_i v_i^*$ (or $F =
\sum_i \sigma_i \ket{u_i}\bra{v_i}$), a sum of $r$ rank-1 outer
products.

An outer product such as $u_i v_i^*$ (or $\ket{u_i}\bra{v_i}$) is
called a one-dimensional \emph{transjector}.  The trans-jector $u_i
v_i^*$ takes a basis vector $v_i \in \CCC^n$ to the corresponding
basis vector $u_i \in \HH$.  By linear superposition, it therefore
takes a general element $x = \sum_j \inner{v_j}{x} v_j \in \CCC^n$ to
$u_i v_i^* x = \inner{v_i}{x} u_i \in \HH$.  Similarly, the adjoint
transjector $v_i u_i^*$ takes $y = \sum_j \inner{u_j}{y} u_j \in \HH$
to $v_i u_i^* y = \inner{u_i}{y} v_i \in \CCC^n$.

The subspace spanned by the $r$ eigenvectors $u_i \in \HH$
corresponding to the $r$ nonzero eigenvalues of $S = F^* F$ will be
denoted as $\U \subseteq \HH$, and the subspace spanned by the $r$
eigenvectors $v_i \in \CCC^n$ corresponding to the $r$ nonzero
eigenvalues of $T = FF^*$ will be denoted as $\V \subseteq \CCC^n$.
The image of $F$ is $\U$, and the image of $F^*$ is $\V$; the kernel
of $F$ is the orthogonal complement $\V^\perp$ of $\V$, and the kernel
of $F^*$ is $\U^\perp$.  $F$ operates by first performing an
orthonormal expansion of $\CCC^n$ using the basis $\{v_i\}$, scaling
each component by $\sigma_i$, and then ``transjecting" to $\U
\subseteq \HH$ by replacing each $v_i$ by the corresponding $u_i$.
$F^*$ similarly ``transjects" from $\HH$ to $\V \subseteq \CCC^n$.

A rank-$r$ matrix $F$ is called an \emph{$r$-dimensional transjector}
if its $r$ nonzero singular values are all equal to 1.
In other words, $F = U Z_r V^*$, where $U$ and $V$ are unitary and
\begin{equation}
\label{eq:zr}
Z_r= \overbrace{\left[
\begin{array}{c|c}
I_r & 0  \\
\hline
0&0 \\
\end{array} \right]}^n.
\end{equation}

Equivalently, $FF^* = U(Z_r Z_r^*)U^* = P_{\U}$ is an $r$-dimensional
orthogonal projector
onto an $r$-dimensional subspace $\U \subseteq \HH$ with an orthonormal
basis
$\{u_i \in \HH, 1 \le i \le r\}$ (the $\U$-basis) consisting of the first
$r$
columns of $U$, and $F^*F = V(Z_r^*Z_r)V^* = P_{\V}$ is an $r$-dimensional
orthogonal  projector onto an $r$-dimensional subspace $\V \subseteq
\CCC^n$ with an 
orthonormal basis
$\{v_i \in \CCC^n, 1 \le i \le r\}$ (the $\V$-basis) consisting of the first
$r$
columns of $V$.  

The SVD $F = U Z_r V^*$ thus reduces to a sum of $r$ one-dimensional
transjectors (outer products):
\begin{equation}
F = \sum_{i=1}^r u_i v_i^*.
\end{equation}
If $u \in \U$, then $u = \sum_{i=1}^r
\inner{u_i}{u}u_i$, and
\begin{equation}
F^*u = \sum_{i=1}^r \inner{u_i}{u}v_i;
\end{equation}
\ie $F^*$ ``transjects" $u$ to a corresponding vector $v
\in \V$.  Similarly, if $v \in \V$, then
\begin{equation}
Fv = \sum_{i=1}^r \inner{v_i}{v}u_i;
\end{equation}
\ie $F$ performs the inverse map from $\V$ to $\U$.  If $u \in \HH$,
then $F^*$ first projects $u$ onto $\U$ and then ``transjects" to $\V$
as above; similarly, for a general $v \in \CCC^{n}$, $F$ first projects
$v$ onto $\V$ and then ``transjects" to $\U$ as above.

An $r$-dimensional transjector $F$ is also called a \emph{partial isometry},
because it is an isometry (distance-preserving transformation) between  the
subspaces $\U \subseteq \HH$ and $\V \subseteq \CCC^{n}$.  Indeed, if $v, v'
\in
\V$ and $u = Fv, u' = Fv'$, then
\begin{equation}
\inner{u}{u'} = u^* u' = v^* F^* F v' = v^* P_{\V} v' = v^* v' =
\inner{v}{v'},
\end{equation}
so inner products and \emph{a fortiori} squared norms and distances are
preserved.  Similarly, if $u, u' \in \U$, then $\inner{F^*u}{F^*u'} =
\inner{u}{u'}$.  However, inner products are not preserved if $u,u'
\notin \U$ or $v,v' \notin \V$.

This discussion is summarized in the following theorem:

\begin{theorem}[Transjectors (partial isometries)]
\label{thm:transjectors}
The following statements are equivalent for a  matrix
$F$ whose columns are $n$ vectors in a complex Hilbert space $\HH$:
\begin{enumerate}
\item $F$ is a transjector (partial isometry) between $r$-dimensional
subspaces $\U \subseteq \HH$ and $\V \subseteq \CCC^{n}$;
\item $FF^* = P_{\U}$ for an $r$-dimensional subspace $\U \subseteq
\HH$;
\item $F^*F = P_{\V}$ for an $r$-dimensional subspace $\V \subseteq
\CCC^{n}$.
\end{enumerate}

A transjector $F$ between $r$-dimensional subspaces $\U \subseteq \HH$ and
$\V
\subseteq \CCC^{n}$ may be expressed as $F = U Z_r V^*$, where $U$ is a
unitary
matrix whose first $r$ columns $\{u_i, 1 \le i \le r\}$ are an orthonormal
basis
for $\U$, $V$ is an $n \times n$ unitary
matrix whose first $r$ columns $\{v_i, 1 \le i \le r\}$ are an orthonormal
basis for $\V$, and  $Z_r$ is given by (\ref{eq:zr}).  Equivalently, $F =
\sum_{i=1}^r u_i v_i^*.$

A transjector $F: \CCC^n \to \U$ (resp.\ $F^*: \HH \to \V$) is an isometry
if
restricted to $\V$ (resp.\ $\U$).
\end{theorem}

\section{Quantum Measurement}
\label{sec:qm}

In this section we present some elements of the theory of quantum
measurement,
following \cite{EF01} and unpublished work in \cite{EF00}.  In the remainder
of
the paper we will develop analogous results for tight frames.

A quantum system in a pure state is characterized by a  normalized vector
$\phi$ in
a Hilbert space $\HH$.  Information about a quantum system is extracted by
subjecting the system to a measurement. In quantum
theory, the outcome of a measurement is inherently probabilistic, with the
probabilities of the outcomes of any conceivable measurement determined by
the
state vector $\phi \in \HH$.

A quantum measurement is described by a collection of Hermitian
operators $\{Q_i\}$ on $\HH$, where the index $i$ corresponds to a possible
measurement outcome.  The laws of quantum mechanics impose certain
mathematical
constraints on the  measurement operators.

In the simplest case, the measurement
operators are rank-one operators and have the outer-product form
$Q_i=\mu_i\mu_i^*$ for some nonzero vectors $\mu_i \in \HH$.
Such
measurements will be called rank-one measurements, and the vectors
$\mu_i$
will be called the measurement vectors.

If the state vector is $\phi$, then the probability
of observing the $i$th outcome is
\begin{equation}
\label{eq:prob}
p(i) = \inner{\phi}{Q_i\phi} = |\inner{\mu_i}{\phi}|^2.
\end{equation}
To ensure that the probabilities $p(i)$ sum to 1 for any
normalized
$\phi \in \HH$, we impose the constraint
\begin{equation}
\label{eq:povm}
\sum_i Q_i  = I_\HH,
\end{equation}
where $I_{\HH}$ is the identity operator on $\HH$;  then
\begin{equation}
\sum_i p(i) = \inner{\phi}{\sum_i Q_i\phi} = \inner{\phi}{\phi} = 1.
\end{equation}
 
We distinguish between standard (von Neumann) measurements and generalized
measurements, or positive operator-valued measures (POVMs).
In a standard measurement, the measurement operators $\{Q_i\}$ form a
complete set of orthogonal projectors. Thus
\begin{eqnarray}
\label{eq:von1}
Q_iQ_i & = & Q_i;  \\
\label{eq:von2}
Q_iQ_j & = & 0, \quad \mbox{if~} i \neq j;  \\
\label{eq:von3}
\sum_i Q_i  & = & I_\HH.
\end{eqnarray}
If the measurement is rank-one, so that $Q_i=\mu_i
\mu_i^*$, then (\ref{eq:von1}) and (\ref{eq:von2}) imply that
$\inner{\mu_i}{\mu_j}=\delta_{ij}$,
while  (\ref{eq:von3}) implies that
\begin{equation}
\label{eq:vonv2}
x=I_\HH x=\sum_i \inner{\mu_i}{x}\mu_i,\,\,\, \forall x \in \HH,
\end{equation}
so the measurement vectors $\{\mu_i\}$ form an orthonormal basis for $\HH$.

Sometimes a generalized measurement is a more efficient way of obtaining
information about the state of a quantum system than a standard measurement
\cite{P90}.  A generalized measurement consists of a set $\{Q_i\}$ of
nonnegative
Hermitian operators, not necessarily projectors, that  satisfy
$\sum_i Q_i  = I_\HH$. Such a set of operators is termed a POVM. If the
measurement
is rank-one so that $Q_i=\mu_i \mu_i^*$, then the measurement
vectors
$\mu_i$ must satisfy
\begin{eqnarray}
\sum_i \mu_i \mu_i^*=I_\HH.
\end{eqnarray}
A POVM is more general than a standard measurement in that the measurement
vectors
$\mu_i$ are not required to be either normalized or orthogonal.

It can be shown that a generalized measurement on a quantum
system can be implemented by introducing an auxiliary system and
performing standard measurements on the combined system. We will
discuss this property in Section~\ref{sec:neumark} in the context of
Neumark's theorem; in Section~\ref{sec:neumarkf} we show that this
property has an analogue for tight frames.

\subsection{Measurement Matrices}
\label{sec:mmatrix}

A rank-one POVM acting on an $r$-dimensional subspace $\U \subseteq
\HH$ in which the system to be measured is known \emph{a priori} to
lie is defined by a set of $n$ measurement vectors $\{\mu_i, 1 \le
i \le n\}$ that satisfy
\begin{equation}
\label{eq:identu}
\sum_{i=1}^n \mu_i \mu_i^* = P_\U,
\end{equation}
\ie the $n$ operators $Q_i=\mu_i \mu_i^*$ must be a resolution of
the identity on $\U$\footnote{Often these operators are supplemented
by a projection $Q_0 = P_{\U^\perp} = I_{\HH} - P_{\U}$ onto the
orthogonal subspace $\U^\perp \subseteq \HH$, so that $\sum_{i=0}^m
Q_i = I_{\HH}$--- \ie the augmented POVM is a resolution of the
identity on $\HH$.}.

The \emph{measurement matrix} $M$ corresponding to a set of
measurement vectors $\mu_i \in \U$ is defined as the matrix of columns
$\mu_i$ \cite{EF01}.  We have immediately from (\ref{eq:identu}) that
\begin{equation}
\label{eq:mm}
MM^*  =  P_\U.
\end{equation}
Thus a matrix $M$ with $n$ columns in $\HH$ is a measurement matrix
for states in the subspace
$\U \subseteq \HH$ if and only if $M$ satisfies (\ref{eq:mm}).

It follows immediately from Theorem \ref{thm:transjectors} that a
measurement
matrix $M$ with $n$ columns in $\HH$ corresponds to a rank-one POVM acting
on an
$r$-dimensional subspace  $\U \subseteq \HH$ if and only if $M$ is a
transjector
(partial isometry) between $\U$ and an $r$-dimensional subspace $\V
\subseteq
\CCC^n$.  Thus $M$ has all the properties enumerated in  Theorem
\ref{thm:transjectors}.

A measurement matrix $M$ represents a standard measurement if and
only if its $n$ columns are orthonormal; \ie if and only if its Gram
matrix satisfies $M^*M =I_n$.  Then $M$ has rank $n$, $\U$ has
dimension $n$, $\V = \CCC^n$, and $M = UZ_nV^*$ for unitary $U$ and $V$,
where $Z_n$ is given by
\begin{equation}
\label{eq:zn}
Z_n=\left[
\begin{array}{c}
I_n\\
\hline
0 \\
\end{array} \right].
\end{equation}

We summarize the properties of measurement matrices in the following
theorem.
\begin{theorem}[Measurement matrices]
\label{thm:measurement}
The following statements are equivalent for a  matrix
$M$ whose columns are $n$ vectors in a complex Hilbert space $\HH$:
\begin{enumerate}
\item $M$ is a measurement matrix corresponding to a rank-one POVM acting on
an
$r$-dimensional subspace $\U \subseteq \HH$;
\item $M$ is a transjector (partial isometry) between $r$-dimensional
subspaces $\U \subseteq \HH$ and $\V \subseteq \CCC^{n}$;
\item $MM^* = P_{\U}$ for an $r$-dimensional subspace $\U \subseteq
\HH$;
\item $M^*M = P_{\V}$ for an $r$-dimensional subspace $\V \subseteq
\CCC^{n}$.
\end{enumerate}

A measurement matrix $M$ corresponding to a rank-one POVM acting on an
$r$-dimensional subspace $\U \subseteq \HH$ may be expressed as
$M = U Z_r V^*$, where $U$ is a unitary matrix whose first $r$ columns
$\{u_i, 1
\le i \le r\}$ are an orthonormal basis for $\U$, $V$ is an $n \times n$
unitary
matrix whose first $r$ columns $\{v_i, 1 \le i \le r\}$ are an orthonormal
basis for $\V$, and  $Z_r$ is given by (\ref{eq:zr}).  Equivalently, $M =
\sum_{i=1}^r u_i v_i^*.$

A measurement matrix $M$ is an isometry if restricted to $\V$.

A measurement matrix $M$ whose columns are $n$ vectors in $\HH$ represents a
standard measurement if and only if its rank is $n$. Then $M = UZ_nV^*$,
where
$Z_n$ is given by (\ref{eq:zn}), and $M^*M = I_n$.
\end{theorem}

\subsection{Neumark's Theorem}
\label{sec:neumark}

Neumark's theorem \cite{P90} guarantees that any POVM with measurement
vectors $\mu_i \in \U$ can be realized by a set of orthonormal
vectors $\tilde{\mu}_i$ in an extended space $\widetilde{\U}$ such
that $\U \subseteq \widetilde{\U}$, so that $\mu_i=P_\U
\tilde{\mu}_i$.

Using the measurement matrix characterization of a POVM and the SVD,
we now obtain a simple statement and proof of Neumark's
theorem. Moreover, our proof is constructive; we explicitly construct
a set of orthogonal measurement vectors such that their projections
onto $\U$ are the original measurement vectors. In
Section~\ref{sec:neumarkf} we use this construction to extend a tight
frame into an orthogonal basis for a larger space.

\begin{theorem}[Neumark's theorem]
\label{thm:neumark}
Let $M$ be a rank-$r$ measurement matrix of an arbitrary POVM, with
$n$ columns in a complex Hilbert space $\HH$.  In other words, $M$ is
a transjector between an $r$-dimensional subspace $\U \subseteq \HH$
and an $r$-dimensional subspace $\V \subseteq \C^n$.  Then there
exists a standard (von Neumann) measurement with measurement matrix
$\widetilde{M}$ which is a transjector between an expanded
$n$-dimensional subspace $\widetilde{\U} \supseteq \U$ in a possibly
expanded complex Hilbert space $\widetilde{\HH} \supseteq \HH$ and
$\C^n$, and whose projection onto $\U$ is $M = P_{\U}\widetilde{M}$.
\end{theorem}
\begin{proof}
Using Theorem \ref{thm:measurement} we may express $M$ as
$M=UZ_rV^*$. Let $u_i$ and $v_i$ denote the columns of $U$ and $V$
respectively.  Assume that $\HH$ is finite-dimensional, and let $k=
\dim~\HH$.

We distinguish between the case $k \geq n$ (\ie $M$ has at least
as many rows as columns), and the case $k < n$
(\ie $M$ has more columns than rows).

In the case $k \geq n$, define $\widetilde{M}=\sum_{i=1}^n u_iv_i^*$;
then $\widetilde{\U} \subseteq \HH$ is the $n$-dimensional subspace
spanned by $\{u_i, 1 \le i \le n\}$.  The projection of
$\widetilde{M}$ onto $\U$ is
\begin{equation}
P_{\U}\widetilde{M} = \sum_{j=1}^m u_j u_j^* \sum_{i=1}^n u_iv_i^* =
\sum_{i=1}^m u_iv_i^* = M.
\end{equation}
Moreover, the columns of $\widetilde{M}$ are orthonormal, since its
Gram matrix is
\begin{equation}
\widetilde{M}^*\widetilde{M}=\sum_{j=1}^n v_ju_j^* \sum_{i=1}^n
u_iv_i^*=\sum_{i=1}^nv_iv_i^*=I_n.
\end{equation}

In the case $k < n$, first embed $\U$ in an $n$-dimensional space
$\widetilde{\U}$ in an expanded complex Hilbert space $\widetilde{\HH}
\supseteq \HH$, and let $\{\tilde{u}_i, 1 \le i \le n\}$ be an
orthonormal basis for $\widetilde{\U}$ of which the first $m$ vectors
are the $\U$-basis.  Then proceed as before, using $\tilde{u}_i$ in
place of $u_i$.
\end{proof}

It is instructive to consider the matrix representation of
$\widetilde{M}$ in both cases. Recall that $M=UZ_rV^*$, where $Z_r$ is
given by (\ref{eq:zr}).

In the case $k \geq n$, we construct $\widetilde{M}$ simply by
extending the identity matrix along the diagonal; thus
$\widetilde{M}=UZ_nV^*$ where $Z_n$ is given by (\ref{eq:zn}).  Thus,
when $k \geq n$, the left and right unitary matrices in the SVD of $M$
and $\widetilde{M}$ are the same, and are equal to $U$ and $V$,
respectively.

If $k = n$, then $Z_n = I_n$ and $\widetilde{M} = UV^*$.

In the case $k < n$, we first replace the left unitary matrix $U$ by
$\widetilde{U}$, and thus replace $k$ by $\tilde{k} = n$; then
$\widetilde{U}$ is an $n \times n$ unitary matrix whose first $r$
columns are the $\U$-basis (where we append $n - k$ zeros to each
basis vector $u_i$). We then define $\widetilde{M} =
\widetilde{U}V^*$.

Examples of the construction of the orthogonal measurement vectors
associated with a given POVM along the lines of this proof will be given in
Section~\ref{sec:neumarkf}, in the context of frames.

\subsection{Optimal Quantum Measurements}
\label{sec:qd}

We now recapitulate some results on optimal quantum measurements
according to various criteria, which will be relevant to the
construction of optimal tight frames.


Let $\{\phi_i,1 \leq i \leq n \}$ be a collection of $n \leq k$
normalized vectors $\phi_i$ in a $k$-dimensional complex Hilbert space
$\HH$, representing different preparations of a quantum system.
In general these vectors are non-orthogonal
and span an \mbox{$r$-dimensional} subspace $\U \subseteq \HH$. The
vectors are linearly independent if $r=n$.

To distinguish between the
different preparations, we subject the system to a measurement.  For
our measurement, we restrict our attention to POVMs consisting of $n$
rank-one operators of the form $Q_i=\mu_i\mu_i^*$ with
measurement vectors $\mu_i \in \U$.  We do not require the vectors
$\mu_i$ to be orthogonal or normalized. However, to constitute a
POVM on $\U$ the measurement vectors must satisfy (\ref{eq:identu}).

If the states are prepared with equal prior probabilities, then the
probability of detection error using the measurement vectors
$\mu_i$ is given from (\ref{eq:prob}) by
\begin{equation}
P_e=1-\frac{1}{n}\sum_{i=1}^n |\inner{\mu_i}{\phi_i}|^2.
\end{equation}
If the vectors $\mu_i$ are orthonormal, then choosing
$\mu_i=\phi_i$ results in $P_e=0$. However, if the given vectors
are not orthonormal, then no measurement can distinguish perfectly
between them.  Therefore, a fundamental problem in quantum mechanics
is to construct measurements optimized to distinguish between a set of
non-orthogonal pure quantum states.

This problem may be formulated as a quantum detection problem, so that
the measurement vectors are chosen to minimize the probability of
detection error, or more generally, minimize the Bayes cost.
Necessary and sufficient conditions for an optimum measurement
minimizing the Bayes cost have been derived \cite{H73,YKL75,H76}.
However, except in some particular cases \cite{H76,CBH89,BKMO97},
obtaining a closed-form analytical expression for the optimal
measurement directly from these conditions is a difficult and unsolved
problem.

An alternative approach proposed in \cite{EF01} is to choose a
different optimality criterion, namely a squared-error criterion, and
to seek measurement vectors that minimize this criterion.  Specifically,
the measurement vectors are chosen to minimize the sum of the squared
norms of the error vectors, where the $i$th error vector is defined as
the difference between the $i$th state vector and the $i$th
measurement vector.  The optimizing measurement is referred to as the
\emph{least-squares measurement} (LSM).

It turns out that the LSM problem has a
simple closed-form solution which has many desirable properties.  Its
construction is relatively simple; it can be determined directly from
the given collection of states; it minimizes the probability of
detection error when the states exhibit certain symmetries
\cite{EF01}; it is ``pretty good'' when the states to be distinguished
are equally likely and almost orthogonal \cite{HW94}; and it is
asymptotically optimal \cite{H96}.

In the next section we will develop a relationship between POVMs and
tight frames. We then apply ideas and results derived in the context
of quantum detection to the construction and characterization of tight
frames.  In particular, we will apply the squared-error criterion developed
in \cite{EF01} to the construction of optimal tight frames.

\newpage 
\section{Tight Frames}
\label{sec:frames}

Frames, which are generalization of bases, were introduced in the
context of nonharmonic Fourier series by Duffin and Schaeffer
\cite{DS52} (see also \cite{Y80}). Recently, the theory of frames has
been expanded \cite{HW89,D90,D92,A95}, in part due to the utility of
frames in analyzing wavelet decompositions.  Here we will focus on tight
frames,
which have particularly nice properties.

Let $\{\varphi_i,1 \leq i \leq n \}$ denote a set of $n$ vectors in an
$r$-dimensional subspace $\U$ of a Hilbert space $\HH$.  The vectors
$\varphi_i$ form a {\em tight frame} for $\U$ if there exists a constant
$\beta > 0$
such that
\begin{equation}
\label{eq:itf}
\sum_{i=1}^n |\inner{x}{\varphi_i}|^2 = \beta^2 ||x||^2,
\end{equation}
for all $x \in \U$ \cite{D92}.  If $\beta = 1$, the tight frame is said to
be
\emph{normalized};  otherwise it is said to be
\emph{$\beta$-scaled}.\footnote{More generally, the vectors $\varphi_i$ form
a 
{\em frame} for $\U$ if there exist constants $\alpha > 0$
and $\beta < \infty$ such that
\begin{eqnarray*}
&& \alpha^2||x||^2 \leq \sum_{i=1}^n |\inner{x}{\varphi_i}|^2 \leq
\beta^2 ||x||^2, 
\end{eqnarray*}
for all $x \in \U$ \cite{D92}.
The lower bound ensures that the vectors $\varphi_i$
span $\U$; thus we must have $n \geq r$. If $n<\infty$, then the right
hand inequality  is always satisfied with
$\beta^2=\sum_{i=1}^n \inner{\varphi_i}{\varphi_i}$. Thus, any finite set of
vectors that spans $\U$ is a frame for $\U$. In particular, any basis
for $\U$ is a frame for $\U$. However, in contrast to basis vectors,
which are linearly independent, frame vectors with $n>r$ are linearly
dependent.  A tight frame is a special case of a frame for which
$\alpha=\beta$.}

Of course any orthonormal basis for $\U$ is a normalized tight frame for
$\U$. 
However, there also exist tight frames for $\U$ with $n > r$, which are
necessarily linearly dependent.
 The \emph{redundancy} of the tight frame is defined as $\rho = n/r$.

Since
\begin{equation}
\label{eq:projid}
\sum_{i=1}^n |\inner{x}{\varphi_i}|^2 = \sum_{i=1}^n x^*\varphi_i
\varphi_i^* x =
\inner{x}{\left(\sum_i \varphi_i \varphi_i^*\right)x},
\end{equation}
the fact that \eqr{eq:itf} holds for all  $x \in \U$ implies that
\begin{equation}
\label{eq:ntf}
\sum_{i=1}^n \varphi_i\varphi_i^* = \beta^2 P_\U.
\end{equation}
Conversely, if the vectors $\varphi_i \in \U$ satisfy
(\ref{eq:ntf}), then \eqr{eq:projid} implies
that (\ref{eq:itf}) is satisfied for all $x \in \U$. We conclude that a set
of $n$
vectors $\varphi_i \in \U$ forms a tight frame for $\U$ if and only if the
vectors
satisfy (\ref{eq:ntf}) for some $\beta>0$.

Comparing (\ref{eq:ntf}) with (\ref{eq:identu}), we conclude that:
\begin{theorem}[Tight frames]
\label{thm:tf}
A set of vectors $\varphi_i \in \U$ forms a $\beta$-scaled tight frame for
$\U$ if 
and only if the scaled vectors $\beta^{-1}\varphi_i$ are the measurement
vectors
of a rank-one POVM on $\U$. In particular, the vectors $\varphi_i$ form a
normalized tight frame for $\U$ if and only if they are the measurement
vectors of
a rank-one POVM on $\U$.
\end{theorem}

 This fundamental relationship between rank-one quantum measurements and
tight
frames will be the basis for the developments in subsequent sections.  In
the next section, we  define frame matrices in analogy to the
measurement matrices of quantum mechanics.  We then use Neumark's
theorem
to extend tight frames to orthogonal bases.  Motivated by the
least-squares measurement of quantum mechanics, in
Section~\ref{sec:optimal} we  address the problem of constructing
optimal tight frames.

\newpage
\subsection{Frame Matrices}
\label{sec:fmatrix}

In analogy to the measurement matrix, we define the {\em frame
matrix} $F$ as the matrix of columns $\varphi_i$, where the vectors
$\varphi_i$ form a tight frame for $\U$. From (\ref{eq:ntf}) it then
follows that
\begin{equation}
\label{eq:ntfm}
FF^* = \beta^2 P_\U.
\end{equation}

The properties of a frame matrix $F$ follow immediately from
Theorem~\ref{thm:tf} and Theorem~\ref{thm:measurement}:
\begin{theorem}[Frame matrices]
\label{thm:frame}
For a  matrix
$F$ whose columns are $n$ vectors in a complex Hilbert space $\HH$ and for a
constant $\beta > 0$, the following statements are equivalent:
\begin{enumerate}
\item $F$ is the frame matrix of a $\beta$-scaled tight frame for an
$r$-dimensional subspace $\U \subseteq \HH$;
\item $\beta^{-1}F$ is a transjector (partial isometry) between
$r$-dimensional subspaces $\U \subseteq \HH$ and $\V \subseteq \CCC^{n}$;
\item $FF^* = \beta^2 P_{\U}$ for an $r$-dimensional subspace $\U \subseteq
\HH$;
\item $F^*F = \beta^2 P_{\V}$ for an $r$-dimensional subspace $\V \subseteq
\CCC^{n}$.
\end{enumerate}

A frame matrix $F$  of a $\beta$-scaled tight frame for an
$r$-dimensional subspace $\U \subseteq \HH$ may be expressed as
$F = \beta U Z_r V^*$, where $U$ is a unitary matrix whose first $r$ columns
$\{u_i, 1 \le i \le r\}$ are an orthonormal basis for $\U$, $V$ is an $n
\times n$
unitary matrix whose first $r$ columns $\{v_i, 1 \le i \le r\}$ are an
orthonormal
basis for $\V$, and  $Z_r$ is given by (\ref{eq:zr}).  Equivalently, $F =
\beta
\sum_{i=1}^r u_i v_i^*.$

A frame matrix $F$  of a $\beta$-scaled tight frame is an isometry if
restricted
to $\V$ and scaled by $\beta^{-1}$.

A frame matrix $F$  of a $\beta$-scaled tight frame whose columns are $n$
vectors
in $\HH$ represents an orthogonal basis for $\U$ (i.e., is an
\emph{orthogonal
frame matrix}) if and only if its rank is
$n$. Then $F = \beta UZ_nV^*$, where
$Z_n$ is given by (\ref{eq:zn}), and $F^*F = \beta^2 I_n$;  i.e., all frame
vectors
have squared norm $\beta^2$.
\end{theorem}

If the vectors $\{\varphi_i,1 \leq i \leq n\}$ form a tight frame for $\U$,
then
any $x \in \U$ may be expressed as a linear combination of these
vectors: $x=\sum_i a_i \varphi_i$. When $n>r$, the coefficients in this
expansion are not unique. A possible choice is
$a_i=\beta^{-2}\inner{\varphi_i}{x}$, because
\begin{equation}
\label{eq:fexp}
\beta^{-2} \sum_{i=1}^n \inner{\varphi_i}{x}\varphi_i = \beta^{-2}FF^*x =
P_{\U}x
= x.
\end{equation}
 This choice of coefficients has the property that among all possible
coefficients
it has the minimal norm \cite{D92,K94}.

The expansion of (\ref{eq:fexp}) is reminiscent of an expansion
of $x$ in terms of an orthonormal basis for $\U$. However, whereas the
vectors in an orthonormal expansion are linearly
independent, the vectors $\varphi_i$ in (\ref{eq:fexp}) are
linearly dependent when $n > r$.

\subsection{Neumark's Theorem and Construction of Tight Frames}
\label{sec:neumarkf}

Neumark's theorem (Theorem~\ref{thm:neumark}) was derived based on the
properties of measurement matrices. Since by Theorem~\ref{thm:tf} frame
matrices
of tight frames have essentially the same properties as measurement matrices
of
rank-one POVMs, we can now obtain an equivalent of
Neumark's theorem for tight frames. The proof is essentially the same
as the proof of Theorem~\ref{thm:neumark}, so we omit it.

\begin{theorem}[Neumark's theorem for tight frames]
\label{thm:neumarkf}
Let $F$ be a rank-$r$ frame matrix, with $n$ columns in a complex
Hilbert space $\HH$ that span an $r$-dimensional subspace $\U
\subseteq \HH$.  Then there exists an orthogonal frame matrix
$\widetilde{F}$ with
equal-norm orthogonal columns that span an expanded $n$-dimensional
subspace $\widetilde{\U} \supseteq \U$ in a possibly expanded complex
Hilbert space $\widetilde{\HH} \supseteq \HH$ such that the projection
$P_{\U}\widetilde{F}$ of $\widetilde{F}$ onto $\U$ is $F$.
\end{theorem}

We remark that given a set of equal-norm orthogonal vectors in
$\widetilde{\U} \supseteq \U$, their projections onto $\U$ will always form
a tight
frame for $\U$ \cite{A95}. Combining this result with
Theorem~\ref{thm:neumarkf},
we can conclude that a set of vectors forms a tight frame for $\U$ if and
only if
the vectors can be expressed as a projection onto $\U$ of a set of
orthogonal vectors with equal norm in a larger space $\widetilde{\U}$
containing $\U$.

Starting with a given frame matrix $F$ in $\U$, the proof of
Theorem~\ref{thm:neumark} gives a concrete construction of an
orthogonal frame matrix $\widetilde{F}$ in
$\widetilde{\U} \supseteq \U$ such that $P_\U
\widetilde{F}=F$.  
We now give two examples of this construction.  We consider first an
example in which $\dim~\HH < n$, and then one in which $\dim~\HH > n$.

\noindent
\textbf{Example 1}.
Consider the four frame vectors
$\varphi_1= [0.35\,\, -0.61]^*$,
$\varphi_2= [0.61\,\, 0.35]^*$,
$\varphi_3= [0.5\,\, -0.5]^*$,  and
$\varphi_4= [0.5\,\, 0.5]^*$.
The frame matrix associated with this frame is
\begin{equation}
\label{eq:Fex}
F=\left[
\begin{array}{rrrr}
0.35 & 0.61 & 0.5 & 0.5 \\
-0.61 & 0.35 & -0.5 & 0.5
\end{array}
\right];
\end{equation}
we may check that $F$ is indeed the frame matrix of a tight frame since
$FF^*=I_2$.

We wish to construct an orthogonal frame matrix $\widetilde{F}$ such
that $F=P_\U \widetilde{F}$. In the proof of Theorem \ref{thm:neumark} for
the case $\dim~\HH < n$, we constructed an $n \times n$ unitary matrix
$\widetilde{F}$ using the SVD $F=U\Sigma V^*$.  Using this construction
here, we
obtain:
\begin{equation}
U=\left[
\begin{array}{rr}
0.5 & -0.87 \\
-0.87 & -0.5
\end{array}
\right],\,\,\,
\Sigma=
\left[
\begin{array}{rrrr}
1 & 0 & 0 & 0\\
0 & 1 & 0 & 0
\end{array}
\right],\,\,\,
V=\left[
\begin{array}{rrrr}
0.70 & 0 &  0.70 & 0 \\
0 & -0.70 & 0 &  -0.70 \\
0.68 & -0.18 & -0.68 & 0.18 \\
-0.18 & -0.68 & 0.18 & 0.68
\end{array}
\right].
\end{equation}
We now define the extended frame matrix $\widetilde{U}$ in accordance
with the proof of Theorem \ref{thm:neumark}.  The first two columns of
$\widetilde{U}$ are uniquely defined as the first two columns of $U$
with zeroes appended. The remaining two columns are arbitrary, as long
as the resulting $\widetilde{U}$ is unitary. A possible choice is:
\begin{equation}
\label{eq:uc}
\widetilde{U}=
\left[
\begin{array}{rrrr}
0.5 & -0.87 &0 & 0\\
-0.87 & -0.5 & 0 & 0\\
0 & 0 & 0.5 & -0.87\\
0 & 0 & -0.87 & -0.5
\end{array}
\right].
\end{equation}
Then
\begin{equation}
\widetilde{F}=\widetilde{U}V^*=
\left[
\begin{array}{rrrr}
0.35 & 0.61 & 0.5 & 0.5 \\
-0.61 & 0.35 & -0.5 & 0.5 \\
0.35 & 0.61 & -0.5 & -0.5 \\
-0.61 & 0.35 & 0.5 & -0.5
\end{array}
\right].
\end{equation}
We may immediately verify that $\widetilde{F}^*\widetilde{F}=I_4$; \ie
$\widetilde{F}$ represents an orthonormal set of vectors.

Since the columns of $F$ span a $2$-dimensional Hilbert space $\U = \HH$,
the projection onto this space is given by
\begin{equation}
P_\U=
\left[
\begin{array}{rrrr}
1 & 0 & 0 & 0 \\
0 & 1 & 0 & 0 \\
0 & 0 & 0 & 0 \\
0 & 0 & 0 & 0
\end{array}
\right],
\end{equation}
and indeed $F=P_\U \widetilde{F}$.

\noindent
\textbf{Example 2}.  We now consider an example in which $\dim~\HH >
n$.  The construction of $\widetilde{F}$ is simpler than in the
previous case because we do not have to extend $\HH$.  Consider the
three frame vectors $\varphi_1= \frac{1}{2}[1\,\, 1\,\, 1]^*$,
$\varphi_2=\frac{1}{2}[-1\,\, 1\,\, 1]^*$, and
$\varphi_3=\frac{1}{2}[\sqrt{2}\,\, 0\,\, 0]^*$.  The frame matrix
associated with this frame is
\begin{equation}
F=\frac{1}{2}\left[
\begin{array}{rrr}
1 & -1 & \sqrt{2}\\
1 & 1 & 0 \\
1 & 1 & 0
\end{array}
\right].
\end{equation}

In order to verify that $F$ is indeed the frame matrix of a tight frame, we
again
determine the SVD
$F=U\Sigma V^*$, which yields
\begin{equation}
U=\left[
\begin{array}{rrr}
0.58 & 0.82 &  0 \\
0.58 & -0.4 & 0.7 \\
0.58 & -0.4 & -0.7
\end{array}
\right],\,\,\,
\Sigma=
\left[
\begin{array}{rrr}
1 & 0 & 0 \\
0 & 1 & 0 \\
0 & 0 & 0
\end{array}
\right],\,\,\,
V=\left[
\begin{array}{rrr}
0.87 & 0 &  0.5 \\
0.29 & -0.82 & -0.5 \\
0.4 & 0.58 & -0.7
\end{array}
\right].
\end{equation}
From Theorem \ref{thm:frame} we conclude that $F$ is indeed the frame
matrix of a tight frame since its nonzero singular values are all equal to
$1$; \ie
$F$ is a transjector. A basis for the subspace $\U$ spanned by the columns
of $F$ is the two vectors
\begin{equation}
u_1=
\left[
\begin{array}{rrr}
0.58 & 0.58& 0.58
\end{array}
\right]^*, \,\,\,
u_2=
\left[
\begin{array}{rrr}
0.82 & -0.4 & -0.4
\end{array}
\right]^*.
\end{equation}
Thus, $P_\U$ is given by
\begin{equation}
P_\U=\sum_{i=1}^2u_iu_i=\left[
\begin{array}{rrr}
1 & 0 & 0 \\
0 & 0.5 & 0.5 \\
0 & 0.5 & 0.5
\end{array}
\right];
\end{equation}
and indeed $FF^*=P_\U$.

We now define an extended frame matrix $\widetilde{F}$ such that
$F=P_\U \widetilde{F}$ and $\widetilde{F}^*\widetilde{F}=I_3$. From
the proof of Theorem \ref{thm:neumark}, we have
\begin{equation}
\widetilde{F}=UZ_3V^*=UV^*=F+u_3v_3^*=
\left[
\begin{array}{rrr}
0.5 & -0.5 &  0.7 \\
0.85 & 0.15 & -0.5 \\
0.15 & 0.85 & 0.5
\end{array}
\right],
\end{equation}
where
\begin{equation}
\label{eq:uvex}
u_3=
\left[
\begin{array}{rrr}
0 & 0.7& -0.7
\end{array}
\right]^*, \,\,\,
v_3=
\left[
\begin{array}{rrr}
0.5 & -0.5 & 0.7
\end{array}
\right]^*.
\end{equation}
Since $P_\U u_3v_3^*=0$, we have immediately that $F=P_\U \widetilde{F}$.

\section{Optimal Tight Frames}
\label{sec:optimal}

It is often of
interest to construct a tight frame from a given set of vectors $\{\phi_i,1
\leq i
\leq n\}$.  Different constructions have been proposed in the literature
\cite{D90,A95,B97}; however, in the general case no optimality properties
are
known for these different constructions.  Using the least-squares
measurement (LSM)
developed in the context of quantum detection \cite{EF01}, we now propose a
systematic method of constructing optimal tight frames from a given set of
vectors.

Thus we seek  to construct a tight frame of vectors
$\{\varphi_i,1 \leq i \leq n\}$ from a given set of vectors $\{\phi_i,1
\leq i \leq n\}$ that span an $r$-dimensional space $\U \subseteq
\HH$.  A reasonable approach is to find a set of vectors $\varphi_i \in
\U$ that are ``closest'' to the vectors $\phi_i$ in the least-squares
sense.  Thus we seek vectors $\varphi_i$ that minimize the squared error
$E$, defined by
\begin{equation}
\label{eq:serror}
E=\sum_{i=1}^n \inner{e_i}{e_i},
\end{equation} 
where $e_i$ denotes the $i$th error vector
\begin{equation}
\label{eq:error}
e_i=\phi_i-\varphi_i,
\end{equation} 
subject to the constraint (\ref{eq:ntf}).

We may wish to constrain
the scaling $\beta$ in (\ref{eq:ntf}), \eg we may seek a normalized
tight frame with $\beta=1$. The optimal frame in this case is derived in
Section~\ref{sec:clsf} and is referred to as the constrained
least-squares frame (CLSF).  Alternatively, we may choose the vectors
$\{\varphi_i \}$ and $\beta$ to satisfy (\ref{eq:ntf}) and to minimize the
squared error $E$ of (\ref{eq:serror}).  The optimal frame is then referred
to as
the unconstrained least-squares frame (ULSF), and is derived in
Section~\ref{sec:ulsf}.

\subsection{Constrained least-squares frame}
\label{sec:clsf}

We first consider the problem of constructing a set of vectors
$\{\varphi_i,1 \leq i \leq n\}$ that minimize $E$ of
\mbox{(\ref{eq:serror})}, subject to the constraint
\begin{equation}
\label{eq:ntfc}
\sum_{i=1}^n \varphi_i\varphi_i^* = \beta_0^2 P_\U,
\end{equation} 
where $\beta_0^2$ is specified. In the case of a normalized tight frame
$\beta_0^2=1$. 

If the vectors $\phi_i$ are mutually orthogonal with
 $\inner{\phi_i}{\phi_i}=\beta_0^2$, then the solution to
 (\ref{eq:serror}) satisfying the constraint (\ref{eq:ntfc}) is simply
 \mbox{$\varphi_i=\phi_i,\,\,1 \leq i\leq n$}, which yields $E=0$.

To derive the solution in the general case, denote by $F$ and $\Phi$
the $k \times n$ matrices whose columns are the vectors $\varphi_i$ and
$\phi_i$, respectively. The squared error $E$ of
(\ref{eq:serror})-(\ref{eq:error}) may then be expressed in terms of
these matrices as
\begin{equation}
\label{eq:errorm}
E=\tr\bl (\Phi-F)^*(\Phi-F) \br=\tr\bl (\Phi-F)(\Phi-F)^* \br.
\end{equation} 
The constraint (\ref{eq:ntf}) may then be restated as
\begin{equation}
\label{eq:constm}
FF^*=\beta_0^2 P_{\U}.
\end{equation}

The least-squares problem of (\ref{eq:errorm}) seeks a frame matrix
$F$ that is ``close'' to the matrix $\Phi$. If the two matrices are
close, then we expect that the underlying linear transformations they
represent will share similar properties.  The SVD of $\Phi$ specifies
orthonormal bases for $\V$ and $\U$ such that the linear
transformations $\Phi$ and $\Phi^*$ map one basis to the other with
appropriate scale factors. Thus, to find an $F$ close to $\Phi$ we
need to find a linear transformation $F$ that performs a map similar
to $\Phi$.  Employing the SVD $\Phi=U\Sigma V^*$, we rewrite the
squared error $E$ of (\ref{eq:errorm}) as
\begin{equation}
\label{eq:errorm2}
E =\tr\bl (\Phi-F)(\Phi-F)^*\br=\tr\bl
U^*(\Phi-F)(\Phi-F)^*U\br=\sum_{i=1}^k \inner{d_i}{d_i},
\end{equation} 
where
\begin{equation}
d_i=(\Phi-F)^*u_i.
\end{equation}
The vectors $\{u_i,\,1 \leq i \leq r\}$ form an orthonormal
basis for $\U$. 
Therefore, the projection operator onto $\U$ is given by
\begin{equation} 
\label{eq:identusvd}
P_{\U} =\sum_{i=1}^r u_i u_i^*.
\end{equation}
Essentially, we want to construct a map $F^*$ such that the images
of the maps defined by $\Phi^*$ and $F^*$  are as close as
possible in the squared norm sense, subject to the  constraint
\begin{equation} 
\label{eq:constmsvd}
FF^* =\beta_0^2 \sum_{i=1}^r u_i u_i^*.
\end{equation}

The SVD of $\Phi^*$ is given by $\Phi^*=V \Sigma^* U^*$. Consequently,
\begin{equation}
\label{eq:lspui}
\Phi^*u_i=
\left\{ 
\begin{array}{ll}
\sigma_i v_i, \hspace{0.1in} & 1 \leq i \leq r; \nonumber \\
0, & r+1 \leq i \leq k,
\end{array}
\right.
\end{equation}
where $0$ denotes the zero vector.  Denoting the image of $u_i$ under
$F^*$ by $a_i=F^* u_i$, for any choice of $F$ satisfying the
constraint (\ref{eq:constmsvd}) we have
\begin{equation}
\label{eq:lsmui}
\inner{a_i}{a_i}=u_i^* FF^* u_i=
\left\{
\begin{array}{ll}
\beta_0^2, \hspace{0.1in}& 1 \leq i \leq r; \nonumber \\
0, & r+1 \leq i \leq k,
\end{array}
\right.
\end{equation}
and 
\begin{equation}
\inner{a_i}{a_j}=u_i^* FF^* u_j=0, \,\,i \neq j.
\end{equation}
Thus the vectors $a_i,\,\,1 \leq i \leq r$, are mutually
orthogonal with $\inner{a_i}{a_i}=\beta_0^2$ and $a_i=0,\,\,m+1
\leq i \leq k$. 
Combining (\ref{eq:lspui}) and (\ref{eq:lsmui}),
we may express $d_i$ as
\begin{equation}
d_i=
\left\{
\begin{array}{ll}
\sigma_i v_i-a_i, \hspace{0.1in}& 1 \leq i \leq r; \nonumber \\
0, & r+1 \leq i \leq k.
\end{array}
\right.
\end{equation}

Our problem therefore reduces to finding a set of $r$ orthogonal
vectors $a_i$ with norm $\beta_0$ that minimize $E=\sum_{i=1}^r
\inner{d_i}{d_i}$, where $d_i=\sigma_i v_i-a_i$. Since the vectors
$v_i$ are orthonormal, the minimizing vectors must be \mbox{$a_i= \beta_0
v_i,\,\,1 \leq i\leq r$}.

Thus the optimal frame matrix $F$, denoted
by $\widehat{F}_c$, satisfies
\begin{equation}
\hmec^* u_i=
\left\{
\begin{array}{ll}
\beta_0v_i, \hspace{0.1in}& 1 \leq i \leq r; \nonumber \\
0, & r+1 \leq i \leq k.
\end{array}
\right.
\end{equation}
Consequently
\begin{equation}
\hmec=\beta_0 \sum_{i=1}^r u_i v_i^*.
\end{equation} 
We may express \hmc in matrix form as
\begin{equation}
\label{eq:lsmm}
\hmec=\beta_0 UZ_r V^*,
\end{equation}
where $Z_r$ is defined by (\ref{eq:zr}).
The residual squared error is then
\begin{equation}
\label{eq:eminc}
E_{min}^c=\sum_{i=1}^r (\beta_0-\sigma_i)^2 \inner{v_i}{v_i}=\sum_{i=1}^m
(\beta_0-\sigma_i)^2.
\end{equation}

Note that if the singular values $\sigma_i$ are distinct, then the
vectors $u_i,\,\,1 \leq i \leq r$ are unique (up to a phase factor
$e^{j\theta_i}$). Given the vectors $u_i$, the vectors $v_i$ are
uniquely determined, so the optimal frame vectors corresponding to
\hmc are unique.  If, on the other hand, there are repeated singular
values, then the corresponding eigenvectors are not
unique. Nonetheless, the choice of singular vectors does not affect
$\hmec$. Indeed, if the vectors corresponding to a repeated singular
values are $\{u_j\}$, then $\sum_j u_j u_j^*$ is a projection onto the
corresponding eigenspace, and therefore is the same regardless of the
choice of the vectors $\{u_j\}$. Thus
\begin{equation}
\sum_j u_j v_j^* =\frac{1}{\sigma} \sum_j
u_ju_j^*\Phi,
\end{equation}
independent of the choice of $\{u_j\}$, and
the optimal frame is unique.

We may express \hmc directly in terms of $\Phi$ as
\begin{equation}
\label{eq:lsmphi}
\hmec=\beta_0 \Phi((\Phi^*\Phi)^{1/2})^{\dagger},
\end{equation}
where $(\cdot)^\dagger$ denotes the Moore-Penrose pseudo-inverse
\cite{GV96}.
Indeed, \mbox{$((\Phi^* \Phi)^{1/2})^\dagger=V((\Sigma^*
\Sigma)^{1/2})^{\dagger} V^*$}, where $((\Sigma^*
\Sigma)^{1/2})^\dagger$ is a diagonal matrix with diagonal elements
$1/\sigma_i$ for $1 \leq i\leq r$ and $0$ otherwise, so that
$\Phi((\Phi^*\Phi)^{1/2})^{\dagger}=UZ_rV^*$.

Alternatively, \hmc may be expressed as
\begin{equation}
\label{eq:lsmphi2}
\hmec=\beta_0 ((\Phi\Phi^*)^{1/2})^\dagger\Phi,
\end{equation}
where 
\mbox{$((\Phi\Phi^*)^{1/2})^\dagger=U((\Sigma
\Sigma^*)^{1/2})^{\dagger} U^*$}.

We note that the optimal frame vectors $\hat{\varphi}_i^c$ satisfy
\begin{equation}
\inner{\hat{\varphi}_i^c}{\phi_i}=[\hmec^* \Phi]_{ii}=\beta_0
[\Phi^*\Phi]^{1/2}_{ii},
\end{equation}
where $[\cdot]_{ii}$ denotes the $ii$th element of the matrix.  This
relation may be used to derive bounds on the inner products
$\inner{\hat{\varphi}_i^c}{\phi_i}$ in terms of the inner products
$\inner{\phi_i}{\phi_j}$; see \cite{H96}.

\subsubsection{Optimal orthogonal basis and the CLSF}
\label{sec:orthogc}

In the previous section, we sought the $\beta_0$-scaled tight frame
that minimizes the least-squares error.  We may similarly seek
the optimal orthogonal vectors with norm $\beta_0$ of the same
form.  We now explore the connection between the resulting optimal
vectors both in the case of linearly independent vectors $\phi_i$
($r=n$), and in the case of linearly dependent vectors ($r<n$).

{\em Linearly independent vectors}: If the vectors $\phi_i$ are
linearly independent and consequently $\Phi$ has full column rank (\ie
$r=n$), then (\ref{eq:lsmphi}) reduces to
\begin{equation}
\hmec=\beta_0 \Phi(\Phi^*\Phi)^{-1/2}.
\end{equation}
The optimal frame vectors $\hat{\varphi}_i^c$ are mutually orthogonal with
equal norm $\beta_0$, since their Gram matrix is
\begin{equation}
\hmec^*\hmec=\beta_0^2
(\Phi^*\Phi)^{-1/2}\Phi^*\Phi(\Phi^*\Phi)^{-1/2}=\beta_0^2 I.
\end{equation}
Thus, the optimal frame is in fact an optimal orthogonal basis for $\U$.

{\em Linearly dependent vectors}: If the vectors $\phi_i$ are linearly
dependent, so that the matrix $\Phi$ does not have full column rank
(\ie $r < n$), then the $n$ frame vectors $\varphi_i$ cannot be mutually
orthogonal since they span an $r$-dimensional subspace.  We now try to
gain some insight into the optimal frame vectors in this case.  Our
problem is to find a set of vectors that are as close as possible to
the $n$ vectors $\phi_i$, which lie in an $r$-dimensional subspace
$\U$.  We now show that these vectors are the projections onto $\U$ of
the set of  norm-$\beta_0$ orthogonal vectors in $\HH$ that are closest
to the vectors $\phi_i$.

To see this, suppose we seek a set of
orthogonal vectors $\tilde{\varphi}_i \in \HH$ with
$\inner{\tilde{\varphi}_i}{\tilde{\varphi}_i}=\beta_0^2$ that are as close
as
possible to the vectors $\phi_i$.  From Theorem~\ref{thm:frame} we
have that
\begin{equation}
\label{eq:tildem}
\sum_{i=1}^n \tilde{\varphi}_i\tilde{\varphi}_i^* =\beta_0^2
P_{\widetilde{\U}},
\end{equation}
where $\widetilde{\U} \supseteq \U$ is the space spanned by the vectors
$\tilde{\varphi}_i$.

Since there are at most $r$ orthogonal vectors in $\U$, imposing an
orthogonality constraint forces the optimal orthogonal vectors
$\tilde{\varphi}_i$ to lie partly in the orthogonal complement $\U^\perp$.
Each vector then has a component in $\U$, $\tilde{\varphi}_i^\U$, and a
component in $\U^\perp$, $\tilde{\varphi}_i^{\U^\perp}$.  Using
(\ref{eq:tildem}), the component in $\U$ satisfies
\begin{equation}
\label{eq:muu}
\sum_{i=1}^n \tilde{\varphi}_i^\U(\tilde{\varphi}_i^\U)^*  =
\sum_{i=1}^n P_\U \tilde{\varphi}_i\tilde{\varphi}_i^*P_\U =\beta_0^2 P_\U
P_{\widetilde{\U}} P_\U=\beta_0^2  P_\U,
\end{equation}
where the last equality follows from the fact that $\U \subseteq
\widetilde{\U}$. 
Now we rewrite the error $E$ of (\ref{eq:serror}) as
\begin{eqnarray}
\label{eq:eorthog}
E & = & \sum_{i=1}^n
\inner{\phi_i-\tilde{\varphi}_i^\U-\tilde{\varphi}_i^{\U^\perp}}{\phi_i
-\tilde{\varphi}_i^\U-\tilde{\varphi}_i^{\U^\perp}} \nonumber
\\
& = & \sum_{i=1}^n\bl
\inner{\phi_i-\tilde{\varphi}_i^\U}{\phi_i-\tilde{\varphi}_i^\U}+
\inner{\tilde{\varphi}_i^{\U^\perp}}{\tilde{\varphi}_i^{\U^\perp}} \br,
\end{eqnarray}
since $\inner{\phi_i-\tilde{\varphi}_i^\U}{\tilde{\varphi}_i^{\U^\perp}}=0$.
From (\ref{eq:muu}) we have that
\begin{eqnarray}
\sum_{i=1}^n 
\inner{\tilde{\varphi}_i^{\U^\perp}}{\tilde{\varphi}_i^{\U^\perp}}
& = & 
\sum_{i=1}^n \inner{\tilde{\varphi}_i}{\tilde{\varphi}_i}-
\sum_{i=1}^n \inner{\tilde{\varphi}_i^\U}{\tilde{\varphi}_i^\U} \nonumber \\
& = & n\beta_0^2- \tr \bl \sum_{i=1}^n
\tilde{\varphi}_i^\U(\tilde{\varphi}_i^\U)^* \br \nonumber \\
& = & n\beta_0^2-\tr(\beta_0^2 P_{\U})=(n-r)\beta_0^2,
\end{eqnarray}
independent of the choice of vectors $\tilde{\varphi}_i$.
Thus, minimization of $E$ is equivalent to minimization of
\begin{equation}
\label{eq:epu}
E'=\sum_{i=1}^n
\inner{\phi_i-\tilde{\varphi}_i^\U}{\phi_i-\tilde{\varphi}_i^\U}.
\end{equation}
Furthermore, from (\ref{eq:muu}) the vectors $\tilde{\varphi}_i^\U$ form a
$\beta_0$-scaled tight frame for $\U$.

We conclude that choosing a set of orthogonal vectors with equal norm
$\beta_0$ that minimize $E$ is equivalent to choosing an optimal
$\beta_0$-scaled 
tight frame for $\U$. The optimal orthogonal
vectors are not unique; however, their projections onto $\U$ are
unique and are just the optimal $\beta_0$-scaled tight
frame vectors.  We may choose the projections
of the optimal orthogonal vectors onto $\U^\perp$ arbitrarily, as long
as the resulting $n$ vectors are orthogonal with norm $\beta_0$.  A
convenient choice is
\begin{equation}
\widehat{\widetilde{F}}_c=\beta_0 \sum_{i=1}^n u_i v_i.
\end{equation}

Indeed, Theorem~\ref{thm:neumarkf} shows that the optimal orthogonal
vectors are just a realization of the optimal frame vectors.  This
theorem guarantees that any $\beta_0$-scaled tight frame may be realized by
a set
of orthogonal vectors with norm $\beta_0$ in an extended space such that
their
projections onto the smaller space are the given frame vectors.  Denoting by
$\hat{\varphi}_i^c$ and
$\hat{\tilde{\varphi}}_i^c$ the optimal frame vectors and orthogonal
vectors, respectively, (\ref{eq:epu}) asserts that
\begin{equation}
\hat{\varphi}_i^c=P_\U \hat{\tilde{\varphi}}_i^c.
\end{equation}

We summarize our results regarding the CLSF in the following theorem:
\begin{theorem}[Constrained least-squares frame (CLSF)]
\label{thm:clsf}
Let $\{ \phi_i \}$ be a set of $n$ vectors in a $k$-dimensional
complex Hilbert space $\HH$ that span an $r$-dimensional subspace $\U
\subseteq \HH$.  Let $\{ \hat{\varphi}_i \}$ denote the optimal $n$ frame
vectors that minimize the least-squares error defined by
(\ref{eq:serror})-(\ref{eq:error}), subject to the constraint
(\ref{eq:ntfc}).  Let $\Phi=U\Sigma V^*$ be the rank-$r$ $k \times n$
matrix whose columns are the vectors $\phi_i$, and let \hmc be the $k
\times n$ frame matrix whose columns are the vectors $\hat{\varphi}_i^c$.
Then the unique optimal \hmc is given by
\[ \hmec=\beta_0 \sum_{i=1}^r u_iv_i^*=\beta_0 UZ_r V^*=
\beta_0 \Phi((\Phi^*\Phi)^{1/2})^{\dagger}=
\beta_0 ((\Phi\Phi^*)^{1/2})^ \dagger\Phi, \]
where $u_i$ and $v_i$ denote the columns of $U$ and $V$
respectively and $Z_r$ is defined by (\ref{eq:zr}).\\
The residual squared error is given by
\[ E_{min}^c=\sum_{i=1}^r (\beta_0-\sigma_i)^2, \]
 where $\{ \sigma_i,\,1 \leq i \leq r
\}$ are the nonzero singular
values of $\Phi$.  \\
In addition,
\begin{enumerate}
\item If $r=n$, 
\begin{enumerate}
\item $\hmec=\beta_0 \Phi(\Phi^*\Phi)^{-1/2}$;
\item $\hmec^*\hmec=\beta_0^2 I_n$, and the corresponding frame vectors are
orthogonal with norm $\beta_0$.
\end{enumerate}
\item If $r<n$,
\begin{enumerate}
\item $\hmec$ may be realized by the $\beta_0$-scaled optimal
orthogonal frame matrix
$\widehat{\widetilde{F}}_c=\beta_0\sum_{i=1}^nu_i v_i^*=\beta_0UZ_nV^*$;
\item the action of the two optimal vector sets in the subspace
$\U$ is the same. 
\end{enumerate}
\end{enumerate}
\end{theorem}  

\subsection{Unconstrained least-squares frame}
\label{sec:ulsf}

We now consider the least-squares problem where the scaling of
the frame is not constrained.  Thus, we seek a set of vectors
$\{\varphi_i\}$ that minimize the squared error $E$ of (\ref{eq:serror}),
subject to
\begin{equation}
\label{eq:ntfu}
\sum_{i=1}^n \varphi_i\varphi_i^* = FF^*=\beta^2 P_\U,
\end{equation} 
where $F$ is the matrix of columns $\varphi_i$, and $\beta>0$.

The derivation of the solution to this minimization problem is very
similar to the derivation of the CLSF of
Section~\ref{sec:clsf}. Following the same steps, we can express $E$
as
\begin{equation}
\label{eq:eint}
E=\sum_{i=1}^k \inner{d_i}{d_i},
\end{equation}
 where $d_i=\sigma_i
v_i-a_i$. 

For any choice of $F$ satisfying the
constraint (\ref{eq:ntfu}) we have
\begin{equation}
\inner{a_i}{a_i}=u_i^* FF^* u_i=
\left\{
\begin{array}{ll}
\beta^2, \hspace{0.1in}& 1 \leq i \leq r; \nonumber \\
0, & r+1 \leq i \leq k,
\end{array}
\right.
\end{equation}
and 
\begin{equation}
\inner{a_i}{a_j}=u_i^* FF^* u_j=0, \,\,i \neq j.
\end{equation}
Thus the vectors $a_i,\,\,1 \leq i \leq r$, are mutually
orthogonal with $\inner{a_i}{a_i}=\beta^2$ and $a_i=0,\,\,r+1
\leq i \leq k$. 

Our problem therefore reduces to finding a set of $r$ orthogonal
vectors $a_i$ with equal norm $\beta$ that minimize (\ref{eq:eint}).
Expressing $E$ as
\begin{equation}
E=\sum_{i=1}^r \bl \sigma_i^2+ \inner{a_i}{a_i}-
2\sigma_i \re{\inner{a_i}{v_i}} \br,
\end{equation} 
where $\re{\cdot}$ denotes the real part, we see that minimization of $E$ is
equivalent to minimization of
\begin{equation}
\label{eq:ep}
E'=r\beta^2-2\sum_{i=1}^r \sigma_i \re{\inner{a_i}{v_i}}=
r\beta^2-2\beta\sum_{i=1}^r \sigma_i \re{\inner{\tilde{a}_i}{v_i}},
\end{equation} 
where $\tilde{a}_i=a_i/\beta$.
To determine the optimal vectors $a_i$ we have to minimize $E'$ with
respect to $\beta$ and $\tilde{a}_i$. Fixing $\tilde{a}_i$ and
minimizing with respect to $\beta$, the optimal value of $\beta$ is given by
\begin{equation}
\label{eq:hatb}
\widehat{\beta}=\frac{1}{r} \sum_{i=1}^r \sigma_i
\re{\inner{\tilde{a}_i}{v_i}}.
\end{equation}
Substituting $\widehat{\beta}$ back into (\ref{eq:ep}), we get that the
vectors
$\tilde{a}_i$ are chosen to maximize
\begin{equation}
\label{eq:epp}
\bl \sum_{i=1}^r \sigma_i \Re \{\inner{\tilde{a}_i}{v_i}\} \br^2,
\end{equation} 
subject to the constraint
\begin{equation}
\label{eq:constt}
\inner{\tilde{a}_i}{\tilde{a}_j}=\delta_{ij}.
\end{equation}
Since the vectors
$v_i$ are 
orthonormal, the minimizing vectors must be \mbox{$\tilde{a}_i=
v_i,\,\,1 \leq i\leq r$}. Substituting into (\ref{eq:hatb}),
\begin{equation}
\label{eq:hatb2}
\widehat{\beta}=\frac{1}{r} \sum_{i=1}^r \sigma_i\re{\inner{v_i}{v_i}}=
\frac{1}{r} \sum_{i=1}^r \sigma_i
\stackrel{\Delta}{=}\alpha,
\end{equation}
and $a_i=\alpha v_i$.

Thus the optimal frame matrix $F$, denoted
by $\widehat{F}_u$, satisfies
\begin{equation}
\hmeu^* u_i=
\left\{
\begin{array}{ll}
\alpha v_i, \hspace{0.1in}& 1 \leq i \leq r; \nonumber \\
0, & r+1 \leq i \leq k.
\end{array}
\right.
\end{equation}
Consequently
\begin{equation}
\hmeu=\alpha \sum_{i=1}^r u_i v_i^*.
\end{equation} 
We may express \hmu in matrix form as
\begin{equation}
\label{eq:lsmmu}
\hmeu=\alpha UZ_r V^*=\alpha\Phi((\Phi^*\Phi)^{1/2})^{\dagger}=\alpha
((\Phi\Phi^*)^{1/2})^\dagger\Phi,
\end{equation}
where $Z_r$ is defined by (\ref{eq:zr}).
The residual squared error is then
\begin{equation}
\label{eq:eminu}
E_{min}^u=\sum_{i=1}^r (\alpha-\sigma_i)^2 \inner{v_i}{v_i}=\sum_{i=1}^r
(\alpha-\sigma_i)^2.
\end{equation}
Recall that $S=\Phi^*\Phi=V \Sigma^* \Sigma V^*$; thus $\tr(S)=\sum_{i=1}^r
\sigma_i^2$. Therefore,
\begin{equation}
\label{eq:emin2}
E_{min}^u=\tr(S)-r\alpha^2.
\end{equation}
Note that as we expect $E_{min}^u \leq E_{min}^c$, where $E_{min}^u$ and
$E_{min}^c$ are given by (\ref{eq:eminu}) and (\ref{eq:eminc})
respectively,  with equality if and only if
$\beta_0=\alpha$.

\subsubsection{Optimal orthogonal basis and the ULSF}
\label{sec:orthogu}

We now explore the connection between the least-squares orthogonal
vectors with unconstrained norm and the ULSF.

{\em Linearly independent vectors}: If the vectors $\phi_i$ are
linearly independent and consequently $\Phi$ has full column rank (\ie
$r=n$), then (\ref{eq:lsmmu}) reduces to
\begin{equation}
\hmeu=\alpha \Phi(\Phi^*\Phi)^{-1/2}.
\end{equation}
The optimal frame vectors $\hat{\varphi}_i^u$ are mutually orthogonal with
equal norm $\alpha^2$,
\begin{equation}
\hmec^*\hmec=\alpha^2
(\Phi^*\Phi)^{-1/2}\Phi^*\Phi(\Phi^*\Phi)^{-1/2}=\alpha^2 I,
\end{equation}
and the optimal frame vectors are  in fact the  optimal orthogonal vectors.

{\em Linearly dependent vectors}: If the vectors $\phi_i$ are linearly
dependent, so that the matrix $\Phi$ does not have full column rank
(\ie $r < n$), then the $n$ frame vectors $\varphi_i$ cannot be mutually
orthogonal since they span an $r$-dimensional subspace.  In analogy to
the constrained case we now show that the optimal orthogonal vectors
are related to the optimal frame vectors through a projection onto the
subspace $\U$, spanned by the vectors $\phi_i$.

Suppose we seek a set of orthogonal vectors
$\tilde{\varphi}_i \in \HH$ with equal norm that
 are as close
as possible to the vectors $\phi_i$. From
Theorem~\ref{thm:frame} we have that
\begin{equation}
\label{eq:tildem2}
\sum_{i=1}^n \tilde{\varphi}_i\tilde{\varphi}_i^* =\beta^2
P_{\widetilde{\U}},
\end{equation}
for some $\beta>0$,
where $\widetilde{\U} \supset \U$ is the space spanned by the vectors
$\tilde{\varphi}_i$.
Now, each  vector $\tilde{\varphi}_i$ has
a component in $\U$, $\tilde{\varphi}_i^\U$, and a component in
$\U^\perp$, $\tilde{\varphi}_i^{\U^\perp}$.
Using (\ref{eq:tildem2}), the component in $\U$ satisfies
\begin{equation}
\label{eq:muu2}
\sum_{i=1}^n \tilde{\varphi}_i^\U(\tilde{\varphi}_i^\U)^*  =
\sum_{i=1}^n P_\U \tilde{\varphi}_i\tilde{\varphi}_i^*P_\U =\beta^2 P_\U
P_{\widetilde{\U}} P_\U=\beta^2  P_\U.
\end{equation}
From (\ref{eq:muu2}) we have that
\begin{eqnarray}
\label{eq:inner}
\sum_{i=1}^n 
\inner{\tilde{\varphi}_i^{\U^\perp}}{\tilde{\varphi}_i^{\U^\perp}}
& = & 
\sum_{i=1}^n \inner{\tilde{\varphi}_i}{\tilde{\varphi}_i}-
\sum_{i=1}^n \inner{\tilde{\varphi}_i^\U}{\tilde{\varphi}_i^\U} \nonumber \\
& = & n\beta^2- \tr \bl \sum_{i=1}^n
\tilde{\varphi}_i^\U(\tilde{\varphi}_i^\U)^* \br \nonumber \\
& = & n\beta^2-\tr(\beta^2 P_\U)=(n-r)\beta^2.
\end{eqnarray}
Rewriting the error $E$ of (\ref{eq:serror}) as in (\ref{eq:eorthog})
and using (\ref{eq:inner}), we conclude that minimization of $E$ is
equivalent to minimization of
\begin{equation}
E'=\sum_{i=1}^n
\inner{\phi_i-\tilde{\varphi}_i^\U}{\phi_i-\tilde{\varphi}_i^\U}+\beta^2(n-r
),
\end{equation}
where from (\ref{eq:muu2}) the vectors $\tilde{\varphi}_i^\U$ form a
$\beta$-scaled
tight frame for $\U$.

Following the derivation of  the ULSF, minimizing $E'$ is
equivalent to minimizing
\begin{equation}
\label{eq:epp2}
E''=r\beta^2-2\beta\sum_{i=1}^r \sigma_i
\re{\inner{\tilde{a}_i}{v_i}}+(n-r)\beta^2=
n\beta^2-2\beta\sum_{i=1}^r \sigma_i
\re{\inner{\tilde{a}_i}{v_i}},
\end{equation}
where $\tilde{a}_i=a_i/\beta$, $a_i=(\widetilde{F}^\U)^* u_i$, and
$\widetilde{F}^\U$ is the matrix of columns $\tilde{\varphi}_i^\U$.
Fixing $\tilde{a}_i$ and
minimizing with respect to $\beta$, the optimal value of $\beta$, denoted by
$\widehat{\widetilde{\beta}}$, is given by
\begin{equation}
\label{eq:hatbo}
\widehat{\widetilde{\beta}}=\frac{1}{n} \sum_{i=1}^r \sigma_i
\re{\inner{\tilde{a}_i}{v_i}}.
\end{equation}
Substituting $\widehat{\widetilde{\beta}}$ back into (\ref{eq:epp2}), we get
that 
the vectors
$\tilde{a}_i$ are chosen to maximize (\ref{eq:epp}) subject to
(\ref{eq:constt}).
Thus, the minimizing vectors are $\tilde{a}_i=
v_i,\,\,1 \leq i\leq r$. Substituting into (\ref{eq:hatbo}) we have
that
\begin{equation}
\label{eq:hatb2o}
\widehat{\widetilde{\beta}}=\frac{1}{n} \sum_{i=1}^r
\sigma_i\re{\inner{v_i}{v_i}}=
\frac{1}{n}\sum_{i=1}^r \sigma_i=\frac{r}{n}\alpha=\frac{\alpha}{\rho},
\end{equation}
where $\alpha$ is defined in (\ref{eq:hatb2}) and $\rho$ is the
redundancy of the frame.
Thus the optimal projections are the columns of
$(1/\rho) \hmeu$,
where $\hmeu$ is the frame matrix of the  ULSF vectors.

We conclude that choosing a set of orthogonal vectors with
unconstrained norm that minimize $E$ is equivalent to choosing an
optimal unconstrained tight frame for $\U$ and scaling these optimal
frame vectors by $1/\rho$.  The optimal unconstrained orthogonal vectors
are not unique; however, their projections onto $\U$ are
unique and are proportional to the optimal
unconstrained tight frame vectors.  We may choose the projections of
the optimal orthogonal vectors onto $\U^\perp$ arbitrarily, as long as
the resulting $n$ vectors are orthogonal with norm $\alpha/\rho$.  A
convenient choice is
\begin{equation}
\widehat{\widetilde{F}}_u=\frac{\alpha}{\rho} \sum_{i=1}^n u_i v_i^*.
\end{equation}

We summarize our results regarding the ULSF in the following theorem:
\begin{theorem}[Unconstrained least-squares frame (ULSF)]
\label{thm:ulsf}
Let $\{ \phi_i \}$ be a set of $n$ vectors in a $k$-dimensional
complex Hilbert space $\HH$ that span an $r$-dimensional subspace $\U
\subseteq \HH$.  Let $\{ \hat{\varphi}_i^u \}$ denote the optimal $n$
frame vectors that minimize the least-squares error defined by
(\ref{eq:serror})-(\ref{eq:error}), subject to the constraint
(\ref{eq:ntfu}).  Let $\Phi=U\Sigma V^*$ be the rank-$m$ $k \times n$
matrix whose columns are the vectors $\phi_i$, and let \hmu be the $k
\times n$ frame matrix whose columns are the vectors $\hat{\varphi}_i^u$.
Then the unique optimal \hmu is given by
\[ \hmeu=\alpha \sum_{i=1}^r u_iv_i^*=\alpha UZ_r V^*=
\alpha \Phi((\Phi^*\Phi)^{1/2})^{\dagger}=
\alpha ((\Phi\Phi^*)^{1/2})^ \dagger\Phi, \]
where $u_i$ and $v_i$ denote the columns of $U$ and $V$
respectively,  $Z_r$ is defined in (\ref{eq:zr}),
$\alpha=\frac{1}{r} \sum_{i=1}^r \sigma_i$ and $\{ \sigma_i,\,1 \leq i \leq
r
\}$ are the nonzero singular
values of $\Phi$.  \\
The residual squared error is given by
\[ E_{min}^u=\sum_{i=1}^r (\alpha-\sigma_i)^2=\tr(\Phi^*\Phi)-r\alpha^2. \]
In addition,
\begin{enumerate}
\item If $r=n$, 
\begin{enumerate}
\item $\hmeu=\alpha \Phi(\Phi^*\Phi)^{-1/2}$;
\item $\hmeu^*\hmeu=\alpha I_n$, and the corresponding frame vectors
are orthogonal with norm $\alpha$.
\end{enumerate}
\item If $r<n$, then
$(1/\rho)\hmeu$ may be realized by the optimal orthogonal frame
matrix
$\widehat{\widetilde{F}}_u=(\alpha/\rho) \sum_{i=1}^n u_i
v_i^*=(\alpha/\rho) UZ_rV^*$.
\end{enumerate}
\end{theorem}

\section{Connection with the Polar Decomposition}
\label{sec:pd}

We now show that the ULSF and the CLSF are related to the polar
decomposition of the matrix $\Phi$.

Let $\Phi$ denote an arbitrary $k \times n$ matrix, where $k \geq n$.
Then $\Phi$ has a {\em polar decomposition} \cite{H86,HJ85},
\begin{equation} 
\label{eq:polar}
\Phi=HY,
\end{equation} 
where $H$ is a $k \times n$ partial isometry
that satisfies $H^*H=I_n$, and
$Y=(\Phi^*\Phi)^{1/2}$. The Hermitian factor $Y$ is always unique; the
partial isometry $H$ is unique if and only if $\Phi$ has full column
rank.  

If $\Phi=U \Sigma V^*$ is the SVD of $\Phi$, then a natural
choice for $H$ is
\begin{equation}
H=UZ_n V^*,
\end{equation}
where $Z_n$ is given by (\ref{eq:zn}). If $r=n$, then this choice of
$H$ is unique. Otherwise $H$ is not unique; however, its projection
onto the column space $\U$ of $\Phi$ is unique
and is given by \cite{E01}
\begin{equation}
\label{eq:ppd}
H_\U =P_\U H=UZ_r V^*=\Phi((\Phi^*\Phi)^{1/2})^{\dagger},
\end{equation}
where $Z_r$ is given by (\ref{eq:zr}).

Comparing (\ref{eq:ppd}) with
(\ref{eq:lsmm}) and (\ref{eq:lsmmu}), we conclude that the ULSF and
CLSF are proportional to the (unique) projection onto $\U$ of the
partial isometry $H$ in a polar decomposition of $\Phi$.  Thus, the ULSF
and CLSF can be computed very efficiently by use of the many known
efficient algorithms for computing the polar decomposition (see \eg
\cite{GV96,ZZ95,H86,D99}).

Recently the truncated polar decomposition (TPD), a variation on the
polar decomposition, has been introduced \cite{BS98} and has proved to be
useful for various estimation and detection problems. As we now show,
the columns of the TPD of a matrix $\Phi$ are just the closest
normalized frame vectors to the columns $\phi_i$ of $\Phi$.

Let $\Phi=U\Sigma V^*$ denote an arbitrary $k \times n$ matrix with
rank $r$. Then the order-$p$ TPD of $\Phi$ is the factorization
\begin{equation}
\label{eq:tpd}
P_{\U_p}\Phi=[U Z_p V^*][V \Sigma^* Z_p V^*]=\tilde{H} \tilde{Y},
\end{equation}
where $P_{\U_p}$ is the orthogonal projection onto the space spanned
by the first $p$ singular vectors $u_i$ of $\Phi$.  From
(\ref{eq:tpd}) it follows that the left-hand matrix in the order-$r$
TPD of $\Phi$ is just the optimal normalized frame matrix $\hmec$.
Similarly, the
left-hand matrix in the order-$p$ TPD of
$\Phi$, with $p<r$, is the optimal normalized tight frame matrix
corresponding to the vectors $P_{\U_p} \phi_i$.

Since the CLSF and ULSF are related to the polar decomposition of
$\Phi$, properties of these optimal frames can be deduced from
properties of the polar decomposition (see \eg
\cite{H86,HJ85,ZZ95,M99}).  For example, the CLSF or ULSF
corresponding to two vector sets $\{\phi_i\}$ and $\{\psi_i\}$ are
the same if and only if the corresponding frame matrices satisfy $\Phi
\Psi^*=|\Phi||\Psi|$, where $|X|=X^*X$ \cite{M99}.

\section{Comparison with Other Proposed Frame Constructions}
\label{sec:compare}

We now compare our results with previously proposed frame
constructions. 

The most popular frame construction from a given set of
vectors is the canonical frame.  Given a set of vectors $\{\phi_i,1
\leq i \leq n\}$ the {\em canonical frame} associated with these
vectors is the frame corresponding to the frame matrix
\cite{D92,B97,BJ99,JB00}
\begin{equation}
\label{eq:canonical}
F=\Phi ((\Phi^* \Phi)^{1/2})^\dagger.
\end{equation}
The canonical frame has many desirable properties. Its construction is
relatively simple; it can be determined directly from the given
vectors; and if the vectors $\phi_i$ are linearly independent, then it
produces an orthonormal basis for $\U$ \cite{D88,HW89,B97}. This
construction was first proposed in the context of wavelets in
\cite{M86}, and plays an important role in wavelet theory
\cite{D88,M89,UA93}.  However, no general optimality properties are known
for
the canonical frame.

Comparing (\ref{eq:canonical}) with (\ref{eq:lsmphi}), we see
immediately that the canonical frame vectors are just the normalized
tight frame vectors that are closest in a least-squares sense to the
vectors $\{\phi_i\}$. Furthermore, the $\beta_0$-scaled tight frame vectors
that
are closest to the vectors $\{\phi_i\}$ are the canonical frame vectors
scaled by
$\beta_0$.  

From Theorem~\ref{thm:ulsf}, it follows that the canonical frame
vectors are the tight frame vectors that minimize the least-squares error
only if
$\alpha=1$, \ie only if $\sum_{i=1}^r \sigma_i=r$. Otherwise, the canonical
frame
is no longer the optimal tight frame in a least-squares sense. However,
if we simply scale each of the canonical frame vectors by $\alpha$,
then the resulting frame minimizes the least-squares error among all
possible tight frames.

\newpage
We summarize our results regarding canonical frames in the
following theorem: 
\begin{theorem}[Canonical frames]
\label{thm:cf}
Let $\{ \phi_i \}$ be a set of $n$ vectors in a $k$-dimensional
complex Hilbert space $\HH$ that span an $r$-dimensional subspace $\U
\subseteq \HH$.  Let $\Phi=U \Sigma V^*$ be the rank-$r$ $k \times n$
matrix whose columns are the vectors $\phi_i$.  Let $u_i$ and $v_i$
denote the columns of the unitary matrices $U$ and $V$ respectively,
let $\{\sigma_i,1 \leq i \leq r \}$ denote the nonzero singular values
of $\Phi$, and let $Z_r$ be defined as in (\ref{eq:zr}).  Let $\{
\varphi_i\}$ be the $n$ canonical frame vectors associated with the
vectors $\phi_i$, and let $F$ denote the matrix of columns
$\varphi_i$. Then
\[F=UZ_rV^*=\Phi((\Phi^*\Phi)^\dagger)^{1/2}=
((\Phi\Phi^*)^{1/2})^\dagger\Phi. \]
In addition,
\begin{enumerate}
\item If $r=n$, 
\begin{enumerate}
\item the canonical frame vectors form an orthonormal basis for $\U$;
\item the canonical frame vectors are the closest orthonormal vectors
to the vectors $\{\phi_i\}$, in a least-squares sense;
\item if $\sum_{i=1}^r \sigma_i=r$, then the canonical frame vectors
are the closest orthogonal vectors with equal norm to the vectors
$\{\phi_i\}$, in a 
least-squares sense;
\item define the scaled canonical frame vectors $\varphi'_i=\beta
\varphi_i$. Then
\begin{enumerate}
\item  the scaled canonical frame vectors are the closest
orthogonal vectors with  norm $\beta$
to the vectors $\{\phi_i\}$, in a least-squares sense;
\item if $\beta= (1/r) \sum_{i=1}^r \sigma_i$, then the scaled
canonical frame vectors are the closest orthogonal vectors with equal
norm to the vectors $\{\phi_i\}$, in a least-squares sense.
\end{enumerate}
\end{enumerate}
\item If $r<n$,
\begin{enumerate}
\item the canonical frame vectors form a tight frame for $\U$;
\item the canonical frame vectors are the closest normalized tight
frame vectors to the vectors $\{\phi_i\}$, in a least-squares sense;
\item if $\sum_{i=1}^r \sigma_i=r$, then the canonical frame vectors
are the closest tight frame vectors to the vectors $\{\phi_i\}$, in a
least-squares sense;
\item Define the scaled canonical frame vectors $\varphi'_i=\beta\mu_i$.
Then
\begin{enumerate}
\item  the scaled canonical frame vectors are the closest
$\beta$-scaled tight frame vectors
to the vectors $\{\phi_i\}$, in a least-squares sense.
\item if $\beta= (1/r) \sum_{i=1}^r \sigma_i$, then the scaled
canonical frame vectors are the closest tight frame vectors
to the vectors $\{\phi_i\}$, in a least-squares sense.
\end{enumerate}
\end{enumerate}
\end{enumerate}
\end{theorem}

\section{Optimal Frames For Geometrically Uniform Vector Sets}
\label{sec:gu}

An important issue in constructing frames from a given set of vectors,
is to what extent the frames inherit the properties of the original
vector set.  In this section we consider the case in which the given
vectors have a strong symmetry property, called geometric uniformity
\cite{F91}.  Under these conditions we can show that the optimal frame has
the same symmetries as the original vector set.

For simplicity, we consider only the optimal normalized tight
frame, which from Theorem~\ref{thm:cf} coincides with the canonical
frame. Since the canonical frame vectors  are proportional to the
vectors constituting the  CLSF and the ULSF, the results extend in a
straightforward manner to these more general constructions.

A set of vectors $\SSS = \{\phi_i, 1 \leq i \leq n\}$ is geometrically
uniform (GU) if every vector in the set has the form $\phi_i=U_i\phi$,
where $\phi$ is an arbitrary vector and the matrices $\{U_i,1 \leq i
\leq n\}$ are unitary and form an abelian group\footnote{That is, $\G$
contains the identity matrix $I$; if $\G$ contains $U_i$, then it also
contains its inverse $U_i^{-1}$; the product $U_i U_j$ of any
two elements of $\G$ is in $\G$; and $U_i U_j=U_j U_i$ for any two
elements in $\G$ \cite{A88}.}  $\G$.

If the vectors $\phi_i$ are GU,
then every row (or column) of the Gram matrix
$S=\{\inner{\phi_i}{\phi_j}\}$ is a permutation of the first row (or
column) \cite{EF01}; such a matrix will be called a permuted matrix.
A set of vectors satisfying
$\inner{\phi_i}{\phi_j}=\inner{\phi_j}{\phi_i}$
 for all $i,j$ (as is the case \eg
for real vector sets) is GU if
and only if the corresponding Gram matrix is a permuted matrix \cite{E01}.

The canonical frame vectors corresponding to a GU vector set are
conveniently characterized in terms of a Fourier matrix defined on an
additive group $G$ isomorphic\footnote{ Two groups $\G$ and $\G'$ are
{\em isomorphic}, denoted by $\G \cong \G'$, if there is a bijection
(one-to-one and onto map) $\varphi: \G \to \G'$ which satisfies
$\varphi(xy)=\varphi(x)\varphi(y)$ for all $x,y \in \G$ \cite{A88}.}
to $\G$.  Specifically, it is well known (see \eg \cite{A88}) that
every finite abelian group $\G$ is isomorphic to a direct product $G$
of a finite number of cyclic groups: $\G \cong G = \Z_{n_1} \times
\cdots \times \Z_{n_p}$, where $\Z_{n_t}$ is the cyclic additive group
of integers modulo $n_t$, and $n = \prod_t n_t$.  Thus every element
$U_i \in \G$ can be associated with an element $g \in G$ of the form
$g = (g_1,g_2,\ldots ,g_p)$, where $g_t \in \Z_{n_t}$; this
correspondence is denoted by $U_i \lra g$.
Each vector $\phi_i=U_i\phi$ is then denoted as $\phi(g)$, where $U_i \lra
g$.

The Fourier transform (FT) of a complex-valued function $\varphi: G
\to \CCC$ defined on $G = \Z_{n_1} \times \cdots \times \Z_{n_p}$ is the
complex-valued function $\hat{\varphi}: G \to \CCC$ defined by
\begin{equation}
\label{eq:fh}
\hat{\varphi}(h) = \frac{1}{\sqrt{n}}\sum_{g \in G} \inner{h}{g}
\varphi(g),
\end{equation}
where the Fourier kernel $\inner{h}{g}$ is
\begin{equation}
\label{eq:gh}
\inner{h}{g} = \prod_{t=1}^p e^{-2 \pi i h_tg_t/n_t}.
\end{equation}
Here $h_t$ and $g_t$ are the $k$th components of $h$ and $g$
respectively, and the product $h_tg_t$ is taken as an ordinary integer
modulo $n_t$.

The FT matrix over $G$ is defined as the $n \times n$ matrix $\FF =
\{\frac{1}{\sqrt{n}}\inner{h}{g}, h,g \in G\}$.  The FT of a column
vector $\varphi = \{\varphi(g), g \in G\}$ is then the column vector
$\hat{\varphi} = \{\hat{\varphi}(h), h \in G\}$ given by
$\hat{\varphi} = \FF\varphi$.  Since $\FF$ is unitary,
we obtain the inverse FT formula
\begin{equation}
\varphi = \FF^*\hat{\varphi} =
\left\{\frac{1}{\sqrt{n}}\sum_{h \in G}
\inner{h}{g}^* \hat{\varphi}(h),g \in G\right\}.
\end{equation}

Following the development in \cite{EF01},
we can now obtain the following
result:
\begin{theorem}[Least-squares normalized tight frames for  GU vector sets]
\label{thm:gu}
Let $\SSS = \{\phi_i = U_i\phi, U_i \in \G\}$, be a geometrically
uniform vector set generated by a finite abelian group $\G$ of unitary
matrices, where $\phi$ is an arbitrary vector, and let $\Phi$ be the matrix
of
columns $\phi_i$. Let $G$ be an additive abelian group isomorphic to $\G$,
let
$\{\phi(g), g \in G\}$ be the elements of $\SSS$ under this isomorphism, and
let 
$\FF$ be the Fourier transform matrix over $G$.
Then the normalized tight frame that is closest in the least-squares sense
to
$\Phi$ is given by the frame matrix
\[
F = \Phi \FF \overline{\Sigma}^\dagger \FF^* = \sum_{h \in
G}u(h)\FF^*(h),
\]
where
\begin{enumerate}
\item $\overline{\Sigma}^\dagger$ is the diagonal matrix whose diagonal
elements
are $\sigma(h)^{-1}$ when $\sigma(h) \neq 0$ and $0$ otherwise,
\item $\{\sigma(h) = n^{1/4}\sqrt{\hat{s}(h)}, h \in G\}$ are
the singular values of $\Phi$,
\item $\{\hat{s}(h), h \in G\}$ is the Fourier transform of the
inner-product
sequence
$\{\inner{\phi(0)}{\phi(g)}, g \in G\}$;
\item $u(h) = \hat{\phi}(h)/\sigma(h)$ when
$\sigma(h) \neq 0$ and $0$ otherwise,
\item $\{\hat{\phi}(h), h \in
G\}$ is the Fourier transform of $\{\phi(g), g \in G\}$;
\item $\FF^*(h)$ is the $h$th row of $\FF^*$.
\end{enumerate}
Finally,
the frame matrix $F$ has the same symmetries as $\Phi$.
\end{theorem}

\subsection{Example of a GU vector set}
We now consider an example demonstrating the ideas of the previous
section.  (The same example was given in \cite{EF01}.) Further
examples and applications of GU vector sets can be found in \cite{EF01}.

Consider the group $\G$ of $n=4$ unitary matrices $U_i$,
where
\begin{equation}
U_1=I_4,\,\,\,
U_2=\left[
\begin{array}{rrrr}
-1 & 0 & 0 & 0 \\
0 & 1 & 0 & 0 \\
0 & 0 & -1 & 0 \\
0 & 0 & 0 & -1
\end{array}
\right],\,\,\,
U_3=\left[
\begin{array}{rrrr}
-1 & 0 & 0 & 0 \\
0 & -1 & 0 & 0 \\
0 & 0 & 1 & 0 \\
0 & 0 & 0 & -1
\end{array}
\right],\,\,\,
U_4=U_2U_3.
\end{equation}
Let the GU vector set be $\SSS=\{\phi_i=U_i\phi,\,\,1 \leq i \leq
4\}$,  where
$\phi=\frac{1}{2}[1\,\, 1\,\, 1\,\, 1]^*$. Then $\Phi$ is
\begin{equation}
\label{eq:phigu}
\Phi=\frac{1}{2}\left[
\begin{array}{rrrr}
1 & -1 & -1 & 1 \\
1 & 1 & -1 & -1 \\
1 & -1 & 1 & -1 \\
1 & -1 & -1 & 1
\end{array}
\right],
\end{equation}
and the Gram matrix $S$ is given by
\begin{equation}
\label{eq:sgu}
S=\frac{1}{2}
\left[
\begin{array}{rrrr}
2 & -1 & -1 & 0 \\
-1 & 2 & 0 & -1 \\
-1 & 0 & 2 & -1 \\
0 & -1 &  -1 & 2
\end{array}
\right].
\end{equation}
Note that the sum of the vectors $\phi_i$ is $0$, so the
vector set is linearly dependent.

In this case $\G$ is isomorphic to $G=\Z_2 \times \Z_2$, \ie
$G=\{(0,0),(0,1),(1,0),(1,1)\}$.
The multiplication table of the group $\G$ is
\begin{equation}
\label{eq:utable}
\begin{array}{c|cccc}
 & U_1 & U_2 & U_3 & U_4 \\
\hline
U_1 & U_1 & U_2 & U_3 & U_4 \\
U_2 & U_2 & U_1 & U_4 & U_3 \\
U_3 & U_3 & U_4 & U_1 & U_2 \\
U_4 & U_4 & U_3 & U_2 & U_1.
\end{array}
\end{equation}
If we define the correspondence
\begin{equation}
\label{eq:coress}
U_1 \lra (0,0),\,\,U_2 \lra (0,1),\,\, U_3 \lra (1,0),\,\,U_4 \lra (1,1),
\end{equation}
then this table becomes the
addition table of $G=\Z_2 \times \Z_2$:
\begin{equation}
\label{eq:atable}
\begin{array}{c|cccc}
 & (0,0) & (0,1) & (1,0) & (1,1) \\
\hline
(0,0) & (0,0) & (0,1) & (1,0) & (1,1) \\
(0,1) & (0,1) & (0,0) & (1,1) & (1,0) \\
(1,0) & (1,0) & (1,1) & (0,0) & (0,1) \\
(1,1) & (1,1) & (1,0) & (0,1) & (0,0).
\end{array}
\end{equation}
Only the way in which the elements are labeled distinguishes the table of
(\ref{eq:atable}) from the table of (\ref{eq:utable});
thus $\G$ is isomorphic to $G$.
Comparing (\ref{eq:utable}) and (\ref{eq:atable})
with (\ref{eq:sgu}), we see that the tables and the matrix $S$ have the
same symmetries.

Over $G=\Z_2 \times \Z_2$, the Fourier matrix $\FF$ is the
Hadamard matrix
\begin{equation}
\FF=\frac{1}{2}\left[
\begin{array}{rrrr}
1 & 1 & 1 & 1 \\
1 & -1 & 1 & -1 \\
1 & 1 & -1 & -1 \\
1 & -1 & -1 & 1
\end{array}
\right].
\end{equation}

Using the equations of the theorem,
we may find the canonical frame:
\begin{equation}
\label{eq:mgun}
F =\frac{1}{2\sqrt{2}}\left[
\begin{array}{rrrr}
1 & -1 & -1 & 1 \\
\sqrt{2} & \sqrt{2} & -\sqrt{2} & -\sqrt{2} \\
\sqrt{2} & -\sqrt{2} & \sqrt{2} & -\sqrt{2} \\
1 & -1 & -1 & 1
\end{array}
\right].
\end{equation}
We verify that the columns $\varphi_i$ of $F$ may be
expressed as $\varphi_i=U_i\varphi_1,\,\,1 \leq i \leq 4$, where
$\varphi_1=\frac{1}{2\sqrt{2}}[1\,\, \sqrt{2}\,\, \sqrt{2}\,\,
1]^*$.  Thus the 
frame vectors $\varphi_i$ also form a GU set generated by $\G$.

\section*{Acknowledgments}

We are grateful to H.~B\"{o}lcskei for encouraging the writing of this
paper. The first author wishes to thank A. V. Oppenheim for his support.

\newpage
\begin{singlespace}
\bibliography{frames}
\bibliographystyle{IEEEbib.bst}
\end{singlespace}

\end{document}